\documentclass[twocolumn]{aastex63}
\usepackage{natbib}
\usepackage{threeparttable}
\usepackage{chngpage}

\newcommand{\squiggle}{SQuIGG$\vec{L}$E \,}
\newcommand{\squigglecomma}{SQuIGG$\vec{L}$E}

\newcommand{\hdelta}{$H_{\delta,A}$ \,}

\newcommand{\re}{$\mathrm{r_e}$ \,}
\newcommand{\recomma}{$\mathrm{r_e}$}

\newcommand{\sersic}{S\'{e}rsic \,}
\newcommand{\sersiccomma}{S\'{e}rsic}

\newcommand{\sigone}{$\mathrm{\Sigma_{1 kpc}} \,$}
\newcommand{\sigonecomma}{$\mathrm{\Sigma_{1 kpc}}$}

\usepackage{amsmath} 

\newcommand{\deltare}{$\mathrm{\Delta log(r_e)}$ \,}
\newcommand{\deltarecomma}{$\mathrm{\Delta log(r_e)}$}

\usepackage{float}
\usepackage{textcomp}

\usepackage{makecell}
\usepackage{totcount}
\newtotcounter{citnum} 
\def\oldbibitem{} \let\oldbibitem=\bibitem
\def\bibitem{\stepcounter{citnum}\oldbibitem}

\date{\today}

\submitjournal{ApJ}

\shorttitle{Compact Structures of Post-Starburst Galaxies}

\shortauthors{Setton et al.}

\begin{document}

\title{The Compact Structures of Massive $z\sim0.7$ Post-Starburst Galaxies in the \squiggle Sample}

\author[0000-0003-4075-7393]{David J. Setton}
\affiliation{Department of Physics and Astronomy and PITT PACC, University of Pittsburgh, Pittsburgh, PA 15260, USA}

\author[0000-0003-1535-4277]{Margaret Verrico}
\affiliation{Department of Physics and Astronomy and PITT PACC, University of Pittsburgh, Pittsburgh, PA 15260, USA}

\author[0000-0001-5063-8254]{Rachel Bezanson}
\affiliation{Department of Physics and Astronomy and PITT PACC, University of Pittsburgh, Pittsburgh, PA 15260, USA}

\author[0000-0002-5612-3427]{Jenny E. Greene}
\affiliation{Department of Astrophysical Sciences, Princeton University, Princeton, NJ 08544, USA}

\author[0000-0002-1714-1905]{Katherine A. Suess}
\affiliation{Department of Astronomy and Astrophysics, University of California, Santa Cruz, 1156 High Street, Santa Cruz, CA 95064 USA}
\affiliation{Kavli Institute for Particle Astrophysics and Cosmology and Department of Physics, Stanford University, Stanford, CA 94305, USA}

\author[0000-0003-4700-663X]{Andy D. Goulding}
\affiliation{Department of Astrophysical Sciences, Princeton University, Princeton, NJ 08544, USA}

\author[0000-0003-3256-5615]{Justin S. Spilker}
\altaffiliation{NHFP Hubble Fellow}
\affiliation{Department of Astronomy, University of Texas at Austin, 2515 Speedway, Stop C1400, Austin, TX 78712, USA}
\affiliation{Department of Physics and Astronomy and George P. and Cynthia Woods Mitchell Institute for Fundamental Physics and Astronomy, Texas A\&M University, 4242 TAMU, College Station, TX 77843-4242}

\author[0000-0002-7613-9872]{Mariska Kriek} 
\affiliation{Leiden Observatory, Leiden University, P.O.Box 9513, NL-2300 AA Leiden, The Netherlands}

\author[0000-0002-1109-1919]{Robert Feldmann}
\affiliation{Institute for Computational Science, University of Zurich, CH-8057 Zurich, Switzerland}

\author[0000-0002-7064-4309]{Desika Narayanan}
\affiliation{Department of Astronomy, University of Florida, 211 Bryant Space Science Center, Gainesville, FL, 32611, USA}
\affiliation{University of Florida Informatics Institute, 432 Newell Drive, CISE Bldg E251 Gainesville, FL, 32611, US}
\affiliation{Cosmic Dawn Centre at the Niels Bohr Institue, University of Copenhagen and DTU-Space, Technical University of Denmark}

\author{Khalil Hall-Hooper}
\affiliation{Department of Mathematics, North Carolina State University, 2108 SAS Hall Box 8205, Raleigh, NC 27695, USA}

\author[0000-0002-0332-177X]{Erin Kado-Fong}
\affiliation{Department of Astrophysical Sciences, Princeton University, Princeton, NJ 08544, USA}

\correspondingauthor{David J. Setton}
\email{davidsetton@pitt.edu}

\begin{abstract}

We present structural measurements of 145 spectroscopically selected intermediate-redshift (z$\sim$0.7), massive ($M_\star \sim 10^{11} \ M_\odot$) post-starburst galaxies from the \squiggle Sample measured using wide-depth Hyper Suprime-Cam i-band imaging. This deep imaging allows us to probe the sizes and structures of these galaxies, which we compare to a control sample of star forming and quiescent galaxies drawn from the LEGA-C Survey. We find that post-starburst galaxies systematically lie $\sim0.1$ dex below the quiescent mass-size (half-light radius) relation, with a scatter of $\sim0.2$ dex. This finding is bolstered by non-parametric measures, such as the Gini coefficient and the concentration, which also reveal these galaxies to have more compact light profiles than both quiescent and star-forming populations at similar mass and redshift. The sizes of post-starburst galaxies show either negative or no correlation with the time since quenching, such that more recently quenched galaxies are larger or similarly sized. This empirical finding disfavors the formation of post-starburst galaxies via a purely central burst of star formation that simultaneously shrinks the galaxy and shuts off star formation. We show that the central densities of post-starburst and quiescent galaxies at this epoch are very similar, in contrast with their effective radii. The structural properties of z$\sim$0.7 post-starburst galaxies match those of quiescent galaxies that formed in the early universe, suggesting that rapid quenching in the present epoch is driven by a similar mechanism to the one at high redshift.


\end{abstract}

\keywords{Post-starburst galaxies (2176), Galaxy quenching (2040), Galaxy evolution (594), Quenched galaxies (2016), Galaxies (573)}

\section{Introduction} \label{sec:intro}

Broadly speaking, galaxies in the Universe can be divided into star-forming and quiescent populations. These populations of galaxies are distinct in that star-forming galaxies form many stars at a rate which is proportional to their stellar mass \citep[e.g.][]{Whitaker2012b}, whereas quiescent galaxies form few or no stars. In addition, the two populations differ structurally at all epochs; star-forming galaxies as a population are systematically larger and less compact than the coeval quiescent population at fixed stellar mass \citep[e.g.][]{VanDerWel2014,Mowla2019,Kawinwanichakij2021}. This indicates that a structural transformation may be coincident with the shutdown of star formation, a process which is jointly referred to as ``quenching". 

There is a growing body of evidence that two distinct pathways to quiescence exist: slow quenching that dominates at low redshift as galaxies gradually exhaust their gas supplies and high redshift rapid quenching that often follows a period of significant starburst \citep{Wu2018, Belli2019, Suess2021}, though there is a significant diversity in quenching times especially among the galaxies which quench more slowly \citep[e.g.][]{Tacchella2022}. The existence of quenched galaxies at high redshift \citep[e.g.][]{Straatman2014, Davidzon2017, Tanaka2019, Forrest2020b, McLeod2021, Valentino2020, DeugenioC2020,Kalita2021} indicates that the seeds of the most massive quiescent galaxies in the local Universe formed on very short timescales through this rapid mode. However, it is still unclear what causes massive galaxies to abruptly quench after an intense period of star formation, and simulations need to invoke various forms feedback to actively suppress star formation and prevent the formation of over-massive galaxies \citep[e.g.][]{Schaye2015, Sijacki2015}. As such, placing empirical constraints on the quenching process, especially in the rapid channel, is essential to understanding the precise process galaxies undergo as they shut off their final epoch of star-forming activity.

Ideally, this process could be studied by finding galaxies at the exact moment preceding the rapid shutdown of their most recent episode of star formation. Unfortunately, rapid shutdown by definition occurs on extremely fast timescales, and it is difficult to identify populations in the midst of shutdown, especially given the uncertainty of future star-formation activity in any galaxy experiencing a starburst at the time of observation. However, it is instead possible to identify the immediate descendants of galaxies which went through the rapid channel by looking for galaxies whose spectra are dominated by a stellar population which formed in the last $<$1 Gyr but which show no evidence of recent star formation. These galaxies are often referred to as post-starburst, or ``K+A" galaxies, due to their composite spectral energy distributions (SEDs) which are dominated by late-type B and A type stars \citep{Dressler1983, Zabludoff1995}. 

Numerous methods have been developed to select post-starburst galaxies, including Balmer absorption strength in conjunction with a measure of weak nebular emission \citep[e.g.][]{Yagi2006,Pracy2005,French2015, Wu2018, Chen2019}, ``K+A" template fitting \citep[e.g.][]{Pattarakijwanich2016}, photometric supercolors \citep[e.g.][]{Wild2014,Almaini2017,Maltby2018,Wilkinson2021}, or UVJ color space \citep[e.g.][]{Belli2019, Suess2020}. All these methods have in common the goal of selecting galaxies with light dominated by a young stellar population with no ongoing star formation \citep[for a detailed review of post-starburst selection methods, see][]{French2021}. Although a dramatic burst of star formation is not strictly required as suggested by the term post-\textit{starburst}--the spectral signatures seen in post-starburst galaxies can be produced by truncation of existing high star formation rate--it is thought that bursts do accompany quenching in high mass galaxies \citep[e.g.][]{French2015,Wild2020, Suess2022}. Post-starburst galaxies serve as a laboratory for understanding the progenitors of the rapid quenching channel and their evolution immediately after star formation shuts off can shed light on the conditions which caused the galaxies to cease forming new stars. 

One important empirical probe of the quenching of massive galaxies comes from the study of their structures. Because star-forming galaxies at any given epoch are consistently larger than quiescent galaxies \citep[e.g.][]{VanDerWel2014,Mowla2019,Kawinwanichakij2021}, it has been suggested that mergers can drive both the quenching and structural transformation of galaxies by driving gas inwards and rendering a more compact aggregate light profile with the resulting centralized star formation \citep{Wellons2015, Zheng2020,Pathak2021}. Mergers appear to be very common in post-starburst galaxies \citep{Zabludoff1996,Pawlik2016,Sazonova2021}, and in simulations mergers have been shown to result in an enhancement in the quiescent fraction in post-merger systems \citep{Quai2021}. A number of structural studies of post-starburst populations have found that they are compact relative to both coeval star-forming and quiescent populations \citep{Yano2016, Almaini2017, Maltby2018, Wu2018, Suess2020} and have younger central stellar populations than their outskirts \citep{Chen2019, DeugenioF2020,Wu2021}. However, the evidence for the ubiquity of merger-driven central starbursts as the driver of structural transformation is inconclusive. The most massive post-starburst galaxies at intermediate-to-high redshift have been found to lack color or age gradients \citep{Maltby2018, Setton2020, Suess2020}, disfavoring out purely nuclear starbursts and implying that star formation prior to quenching may have occurred on kiloparsec scales. These field studies are in contrast with observations of post-starburst galaxies in dense clusters, where environmental effects such as ram-pressure stripping have been shown to be the dominant quenching mechanisms \citep[e.g.][]{Matharu2020, Matharu2021, Werle2022}.

In order to use structures to understand the quenching process, one would ideally like to track the evolution of post-starburst galaxies as a function of their time since quenching. However, precise timing of the time since quenching requires high quality rest-frame optical spectra. Because the post-starburst population does not emerge significantly until $z\sim1$ \citep{Whitaker2012a, Wild2016}, those spectra must be obtained at a minimum of intermediate-redshift to catch even the tail end of the rapid quenching that dominates in the early universe. To date, spectroscopic studies of intermediate-redshift post-starburst galaxies have been limited to small samples \citep[e.g.][]{Wu2018,Wild2020}. However, spectroscopic samples of post-starburst galaxies at intermediate-redshift are accessible in the Sloan Digital Sky Survey (SDSS) thanks in large part to the CMASS BOSS sample \citep{Dawson2013} which targeted high-mass red galaxies at intermediate-redshift. Leveraging this spectroscopic sample in addition to a handful of ancillary SDSS programs, we have launched the \squiggle Survey \citep{Suess2022}, which spectroscopically identified 1318 post-starburst galaxies in the SDSS at $z>0.5$ with spectral signatures that indicate that they have recently shut off their primary epoch of star formation. Crucially, these spectra allow us to characterize the properties of the burst, to measure the time since star formation quenched, and to track the evolution of the sample in key properties as a function of the time since quenching. This sample has allowed us to study in detail the link between the star formation shutdown and AGN incidence \citep{Greene2020}, molecular gas content \citep{Suess2017, Bezanson2022}, and the incidence of mergers (M. Verrico et al. 2021 in preparation).

In this work and its companion letter (Verrico et al. 2021 in preparation), we match 145 post-starburst galaxies in the \squiggle Survey to deep imaging in the Hyper Suprime-Cam Survey \citep{Aihara2018, Aihara2022} in order to study their sizes, structures, and merger signatures and connect these properties to those of the star formation histories as derived from the SDSS spectroscopy. In Section \ref{sec:data}, we introduce the \squiggle sample and a mass-matched control sample of quiescent and star-forming galaxies to which we compare. In Section \ref{sec:sizes}, we detail our methodology for measuring sizes and structures. In Section \ref{sec:structure}, we present our analysis of the sizes and structures. Finally, in Section \ref{sec:discussion}, we discuss our results and their significance to understanding the rapid quenching pathway of galaxy evolution. Throughout this paper we assume a concordance $\Lambda$CDM cosmology with $\Omega_{\Lambda}=0.7$, $\Omega_m=0.3$ and $H_0=70$ $\mathrm{km\,s^{-1}\,Mpc^{-1}}$, and quote AB magnitudes. All reported values of the effective radius (\recomma) are measurements of the semi-major axis and are not circularized.

\begin{figure}
\includegraphics[width=0.45\textwidth]{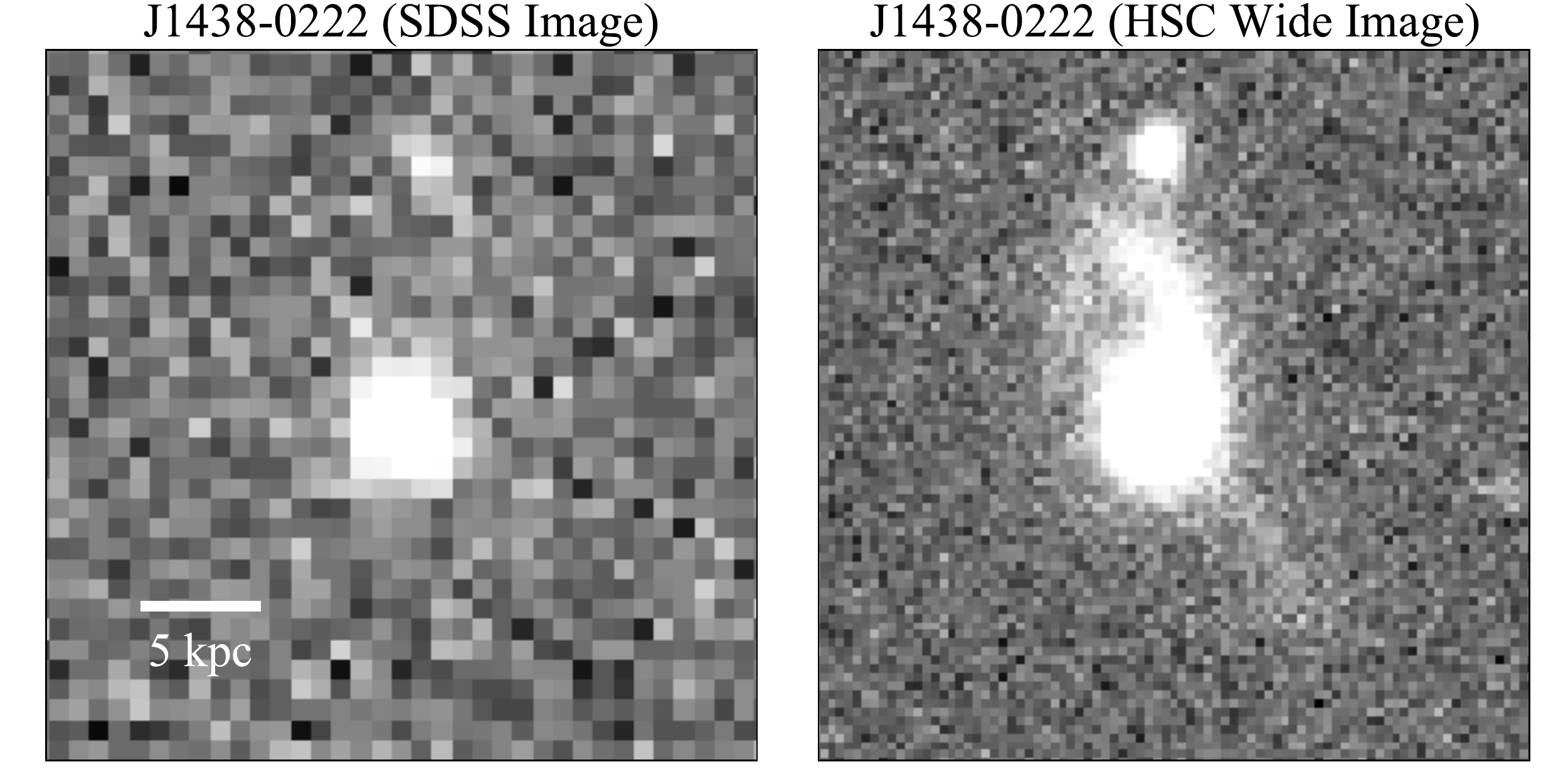}
\caption{i-band images of J1438-0222, a z=0.698 post-starburst galaxy from the SDSS (left) and the Hyper Suprime-Cam Wide Survey. The cutouts are centered at the same physical location, and the pixel scales are 0.396 and 0.168 "/pixel respectively. The combination of better resolution and deeper images resolves faint, low surface brightness light that was previously inaccessible at this redshift, allowing us to accurately measure sizes and identify merger signatures.
\label{fig:sdss_v_hsc}}
\end{figure}

\section{Data} \label{sec:data}

\subsection{The \squiggle Sample} \label{subsec:squiggle}

In order to study the descendants of the rapid quenching process, we turn to the \squiggle Survey. The \squiggle Sample is selected from the Sloan Digital Sky Survey DR14 spectroscopic sample \citep{Abolfathi2018} using the rest-frame $U_m$, $B_m$, and $V_m$ medium-band filters from \cite{Kriek2010} to select galaxies with strong Balmer breaks (indicated by red $U_m-B_m$ colors) and blue slopes redward of the break (indicated by blue $B_m-V_m$ colors). In \cite{Suess2022}, we show that this method reliably selects post-starburst galaxies which quenched within the last $\sim500$ Myr, 75\% of which formed $>25\%$ of their stellar mass in a recent burst, and further discuss the specific selection effects and how they differ from other post-starburst selection techniques. While SDSS imaging exists for the entire sample, the depth and resolution is not sufficient to resolve the galaxies. However, the Hyper Suprime-Cam (HSC) Wide Imaging Survey \citep{Aihara2018} overlaps with $\sim10\%$ of the galaxies in \squiggle and is deep enough to resolve the main galaxies in addition to faint structures, as seen in Figure \ref{fig:sdss_v_hsc}. 

\begin{figure*}[ht!]
\includegraphics[width=0.49\textwidth]{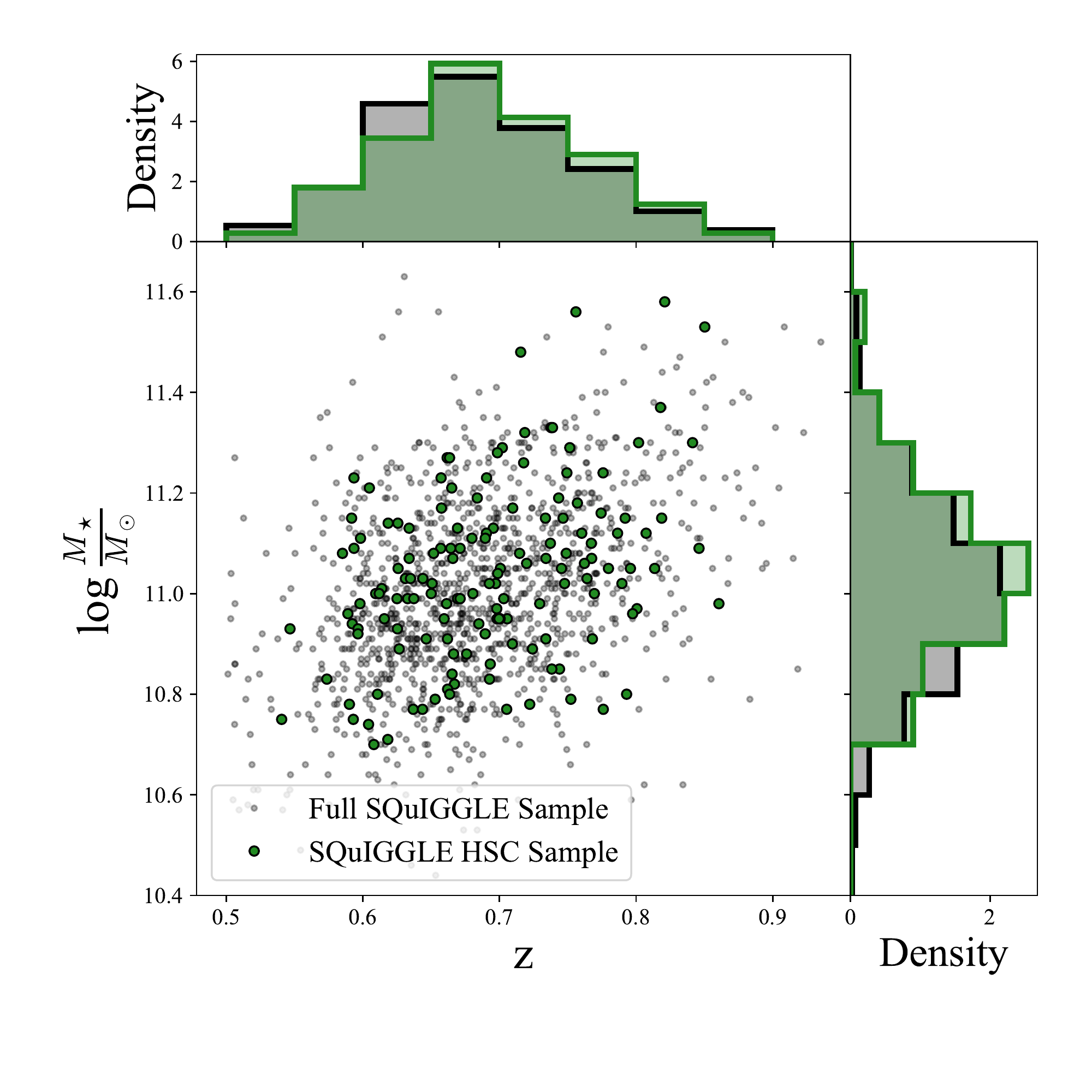}
\includegraphics[width=0.49\textwidth]{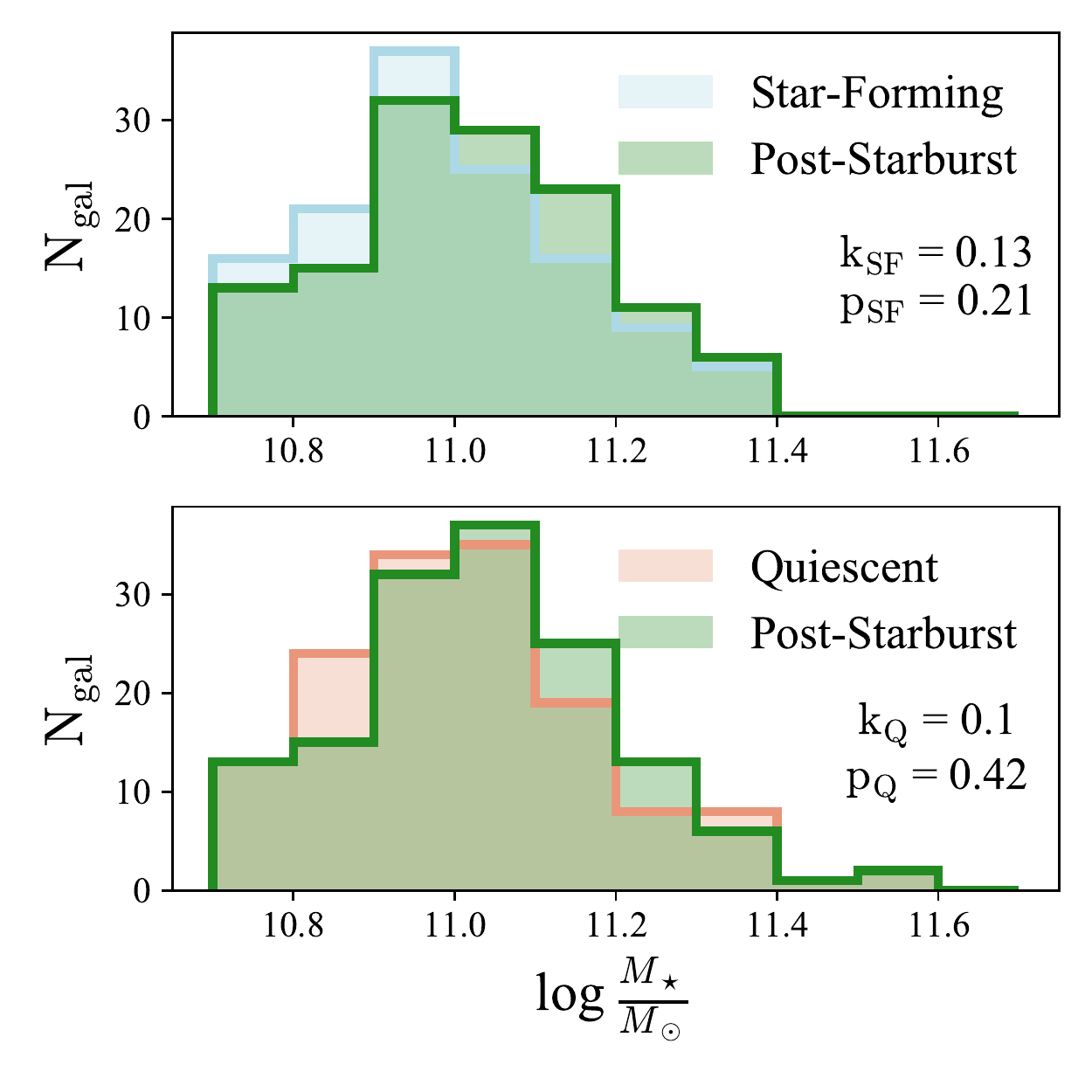}
\caption{(Left): Stellar mass versus redshift for the full \squiggle post-starburst sample (grey) and the HSC sample in this work (green). Above the mass-cut of $M_\star>10^{10.7} \ M_\odot$, the HSC sample spans the full range of \squigglecomma. (Right): The stellar-mass distributions for the mass-matched far-forming (blue) and quiescent (red) matches with the \squiggle sample. The samples are well matched in stellar mass, as indicated by the results of Kolmogorov–Smirnov tests on the two histograms.
\label{fig:mass_redshift}}
\end{figure*}

In order to characterize the stellar populations of these galaxies, we perform two sets of spectral energy distribution (SED) modelling. The first, described in \cite{Setton2020}, are performed on the the SDSS spectra and \textit{ugriz} photometry using \texttt{FAST++}\footnote{\href{https://github.com/cschreib/fastpp}{https://github.com/cschreib/fastpp}}, an implementation of the \texttt{FAST} stellar population fitting code \citep{Kriek2009}. We assume a delayed exponential star formation history (SFR $\sim t e^{-t / \tau}$), BC03 stellar population libraries \citep{Bruzual2003}, a \citet{Chabrier2003} initial mass function, and a \citet{Calzetti1997} dust law. While these fits are limited to capturing only the recent burst of star formation in post-starburst galaxies due to the imposition of a single rise and fall in the star formation history, the physical measurements derived from this common parameterization can easily be compared to other galaxies in the literature which are fit similarly. 

In addition to the traditional parametric fits, we perform non-parametric fits using a custom implementation of \texttt{Prospector} \citep[][see \citealp{Suess2022} for details]{Johnson2017, Leja2017, Johnson2021}. These star formation histories impose their own set of priors on the stellar mass which result in higher stellar masses than those derived with a delayed exponential star formation history \citep[e.g.][]{Lower2020,Leja2021}. In addition, non-parametric star formation histories are still only weakly constraining on star formation before the recent burst due to the outshining from the recently formed stellar population; as such, the prior chosen for early formation times significantly affects the conclusions about the burst mass fraction. However, these non-parametric fits are very useful for providing lower limits on the burst mass fraction with the conservative priors chosen, for robustly recovering the time since quenching, and for reliably measuring instantaneous star formation rates. For this work, we use the stellar masses derived from the delayed exponential star formation history for consistency in comparisons to other mass-size relations, and employ the non-parametric star formation histories to investigate trends within the \squiggle post-starburst sample. 

\subsection{Coeval control sample from the LEGA-C Survey} \label{subsec:legac}

In order to contextualize the sizes and structures of \squiggle post-starburst galaxies relative to coeval star-forming and quiescent galaxies, we require a sample of $z\sim0.7$ galaxies with high quality spectra which fully overlaps with HSC. For this, we turn to the LEGA-C Survey DR3 \citep{VanDerWel2021}. The LEGA-C Survey consists of deep ($\sim20$ hours/galaxy) spectroscopy of galaxies in the COSMOS field. Because the \squiggle sample was selected from a variety of SDSS target selections with a variety of different color and magnitude cuts, we do not yet have a full understanding of how mass complete or representative the sample of post-starburst galaxies are. As such, we elect to compare to a very conservative subset of LEGA-C galaxies above which completeness correction factors are negligible to ensure that the control population we are comparing to is fully representative of the galaxy population at $z\sim0.7$. In this work, we use the sub-sample of LEGA-C galaxies with stellar masses (derived similarly to the \squiggle sample using delayed-exponential star formation histories) above $10^{10.7} \ M_\odot$, slightly higher than the characteristic mass of the sample and well into the regime where completeness corrections are $\sim1$ \citep{VanDerWel2016}. We divide the LEGA-C sample into quiescent and star-forming by their UVJ colors as in \cite{VanDerWel2016} and use those samples as a control group to compare morphology and size measures to the post-starburst galaxies in \squigglecomma, using similarly derived delayed exponential star formation history masses. 

The LEGA-C sample was selected using a redshift dependent K-band magnitude cut, in contrast with \squigglecomma, which was selected using rest-frame colors and a signal-to-noise cut from the whole of SDSS, largely from luminous red galaxies targeted using the CMASS selection criteria \citep{Dawson2013}. As such, the mass and redshift distributions of the two samples are different, even above the previously discussed stellar-mass threshold. In order to fairly compare structural measurements, we create a mass-matched subsample of LEGA-C following the procedure described in detail in M. Verrico et al. in preparation. Briefly, for each galaxy in \squiggle we select the quiescent and star-forming galaxy in LEGA-C which is within 0.05 dex in stellar mass and is the closest in redshift to our main galaxy without replacement.

We perform this procedure on the full sample of \squiggle galaxies with high quality HSC imaging (see Section \ref{subsec:HSC}) resulting in a final sample of 144 mass-matched pairs of post-starburst and quiescent and 129 pairs of post-starburst and star-forming galaxies. For the fifteen most massive galaxies in the \squiggle sample with high quality HSC imaging, there is no similar-mass star-forming counterpart. This dearth of massive star-forming galaxies could be expected in a pencil-beam survey like LEGA-C given the steep high-mass end of the star-forming stellar mass function \citep[e.g.][]{Muzzin2013}. In addition, we are not able to match one of the most massive post-starburst galaxies to a quiescent counterpart. The results of the mass-matching procedure are shown as histograms in Figure \ref{fig:mass_redshift}b. In both cases, a KS test finds that the stellar mass distributions of the post-starburst and control samples are consistent with being drawn from the same distribution ($\mathrm{k_{Q}} = 0.10, \ \mathrm{p_Q} = 0.42$; $\mathrm{k_{SF}} = 0.13, \ \mathrm{p_{SF}} = 0.21$). For testing of size-fitting methodologies and the determination of the mass-size relations for star-forming and quiescent samples, we utilize the entirety of LEGA-C's mass-representative ($\mathrm{log}\frac{M_\star}{M_\odot}>10.7$) sample. For all other comparisons, we utilize the mass-matched samples, and show the star-forming matched post-starburst sample as dashed histograms for comparison with the star-forming disxtributions. We note that by construction, the redshift distributions of the mass-matched samples, which were a secondary priority, overlap less precisely than those of the stellar masses. The LEGA-C quiescent and star-forming samples are a median 0.038 and 0.058 higher in redshift than \squigglecomma. We consider this offset to be acceptable for the secondary parameter, as we are primarily interested in structural parameters of galaxies as they relate to stellar mass and the effects of surface brightness dimming are small over this narrow range in redshift. 

In addition to fully overlapping with HSC, the LEGA-C Survey has the added bonus of being fully observed with existing HST/ACS F814 imaging. This allows us to also use the galaxies in LEGA-C as a test to ensure that the inferences we make using ground-based data agree with the higher-resolution space-based data afforded by HST. 

\begin{figure*}
\includegraphics[width=\textwidth]{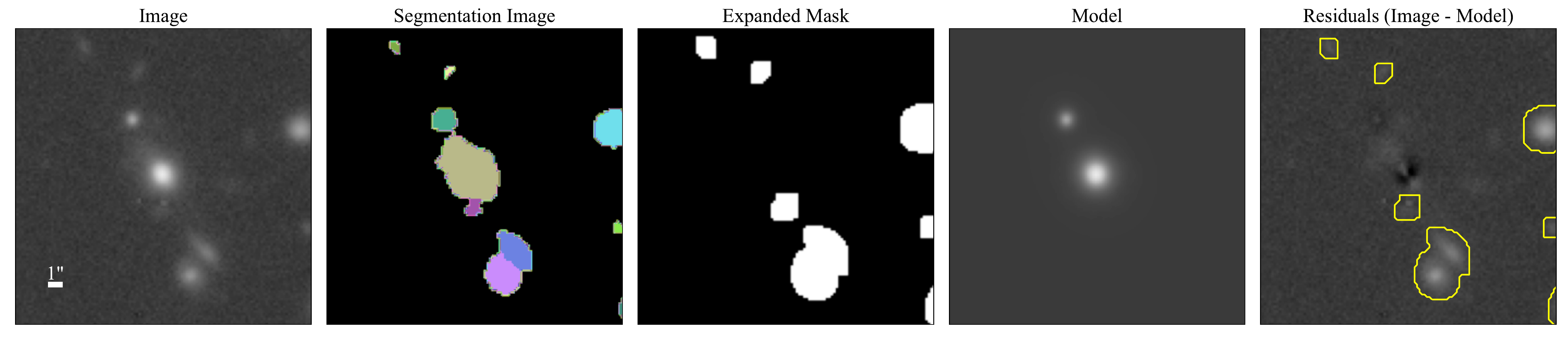}
\caption{A demonstration of the multi-component fitting procedure on a \squiggle post-starburst galaxy. We first show the unmasked image centered on the galaxy of interest in a 20"x20" cutout, with nearby interlopers. The second panel demonstrates the source identification procedure described in Section \ref{subsec:sersic}, which successfully detects and deblends all sources in the field of view. The third panel shows the mask that we provide to \texttt{GALFIT}, which is the segmentation map for all sources $>$25 pixels from the galaxy of interest or which are within 25 pixels but are 3 dex fainter than the main source convolved with a 3-pixel tophat kernel. Note that the object to the Northeast of the galaxy of interest is bright and close enough to not be masked. In the fourth panel, we show the best fitting model produced by \texttt{GALFIT}, and in the final panel we show the residuals on the same color scale as the image and the model, with the masked regions outlined. The fitting successfully accounts for the majority of the light from both the main galaxy and the neighbor.}
\label{fig:sersic_demonstration}
\end{figure*}

\subsection{Hyper-Suprime Cam Imaging \label{subsec:HSC}}

In Figure \ref{fig:sdss_v_hsc}, we show that the SDSS images of \squiggle galaxies are insufficient in both depth and resolution to study the structures of post-starburst galaxies at $z\sim0.7$. However, imaging from the Hyper-Surprime Cam (HSC) Survey \citep{Aihara2018} can be used to robustly obtain sizes from the ground out to $z<1$ \citep{Kawinwanichakij2021}. As of PDR3, the HSC Wide Survey has taken data in at least one \textit{grizy} band \citep{Kawanomoto2018} of $\sim1300 \ \mathrm{deg^2}$ of the sky at depths $\sim3$ orders of magnitude deeper than the SDSS \citep{Aihara2022}, making it ideal for the detailed study of \squiggle post-starburst galaxy images. The survey design is such that the i-band was observed to great depths (26.2 mag point source limit), and, perhaps more importantly, at extremely high resolution (point spread function FWHM $\sim0.6$", a factor of $\sim2$ improvement over SDSS). As such, we elect to perform all our analysis on the i-band images which provide the best combination of depth and seeing and overlap with $\sim10\%$ of the \squiggle sample. 

For every galaxy in \squiggle with i-band imaging in the HSC footprint, we pull a 48x48" coadd cutout and the corresponding point spread function (PSF) model from the PSF Picker\footnote{\href{https://hsc-release.mtk.nao.ac.jp/psf/pdr3/}{https://hsc-release.mtk.nao.ac.jp/psf/pdr3/}} \citep{Bosch2018}. After performing the same mass-cut ($\mathrm{log}\frac{M_\star}{M_\odot}>10.7$) as on LEGA-C, we find that 150 \squiggle galaxies have been imaged in the HSC i-band. We visually inspect all galaxies and exclude images with clear visual artifacts (e.g. cosmic rays, image streaks) in the region of the central object, as well as objects with extremely bright nearby sources which significantly alter the sky subtraction. This removes 5/150 galaxies with HSC coverage, leaving us with a total of 145 post-starburst galaxies with clean images. A full gallery of the \squiggle sample cutouts is shown in Verrico et al. in preparation. The results of this selection in the mass-redshift plane are shown in Figure \ref{fig:mass_redshift}. The subset of \squiggle galaxies we study in this work spans the range of the full \squiggle sample above the mass-completeness cut. 

In addition, we pull cutouts using the identical procedure as above for the entirety of the LEGA-C mass-representative sample. The HSC Survey consists of 3 sky layers: wide, deep, and ultra deep, covering 1400 
$\mathrm{deg}^2$, 26 $\mathrm{deg}^2$, and 3.5 $\mathrm{deg}^2$ respectively and each $\sim1$ dex deeper than the shallower layer \citep{Aihara2018}. Every galaxy in \squiggle was observed only at wide depth; however, the LEGA-C field overlaps with the deep layer of HSC. In order to facilitate a fair comparison to the \squiggle sample, we utilize the wide depth reductions of the LEGA-C galaxies to ensure that the surface brightness limits are similar between the two samples. However, we also pull the full depth HSC Deep cutouts and PSF models for the entire sample, which we use to test the reliability of our fitting procedures for deeper images.


\subsection{Tidal Feature Classifications}

In M. Verrico et al. in preparation, we present visual classifications of the incidence of tidal features in the mass-matched \squiggle and coeval LEGA-C samples. In brief, we instructed eleven members of the \squiggle team and the Pitt Galaxy Group to assign a binary classification of ``Tidally Disturbed" or ``Not Disturbed" to postage stamps of the \squiggle and mass-matched LEGA-C control samples in random order. We divide galaxies into 3 categories: disturbed ($>70\%$ agreement that a tidal feature is present, $\mathrm{N_{gal}}=61$), non-disturbed ($>70\%$ agreement that no tidal features are present, $\mathrm{N_{gal}=51}$), and ambiguous (all galaxies which meet neither of these conditions, often due to the presence of neighbors where the association is unclear and rankers were divided, $\mathrm{N_{gal}}=33$). Throughout this work, we utilize the disturbed and non-disturbed sub-samples of \squiggle to test whether structural parameters vary based on whether or not a galaxy has clear tidal features. 


\section{Galaxy size and structure fitting} \label{sec:sizes}

\subsection{Measuring \sersic sizes} \label{subsec:sersic}

We utilize \texttt{GALFIT} \citep{Peng2010} to quantify the structural parameters of \squiggle post starburst galaxies and the LEGA-C control sample. For each image, we identify and deblend all light sources detected at the 5$\sigma$ level above the background using the Python \texttt{astropy photutils} package \citep{astropy2013,astropy2018}, using the following settings which were chosen to optimally deblend based on visual inspection of the resultant segmentation maps:

\begin{itemize}
    \item[] \texttt{fwhm\_smooth} = 3
    \item[] \texttt{sigma\_detect} = 3
    \item[] \texttt{npixels} = 5
    \item[] \texttt{npixels\_deblend} = 5
\end{itemize}

\begin{figure*}
\includegraphics[width=\textwidth]{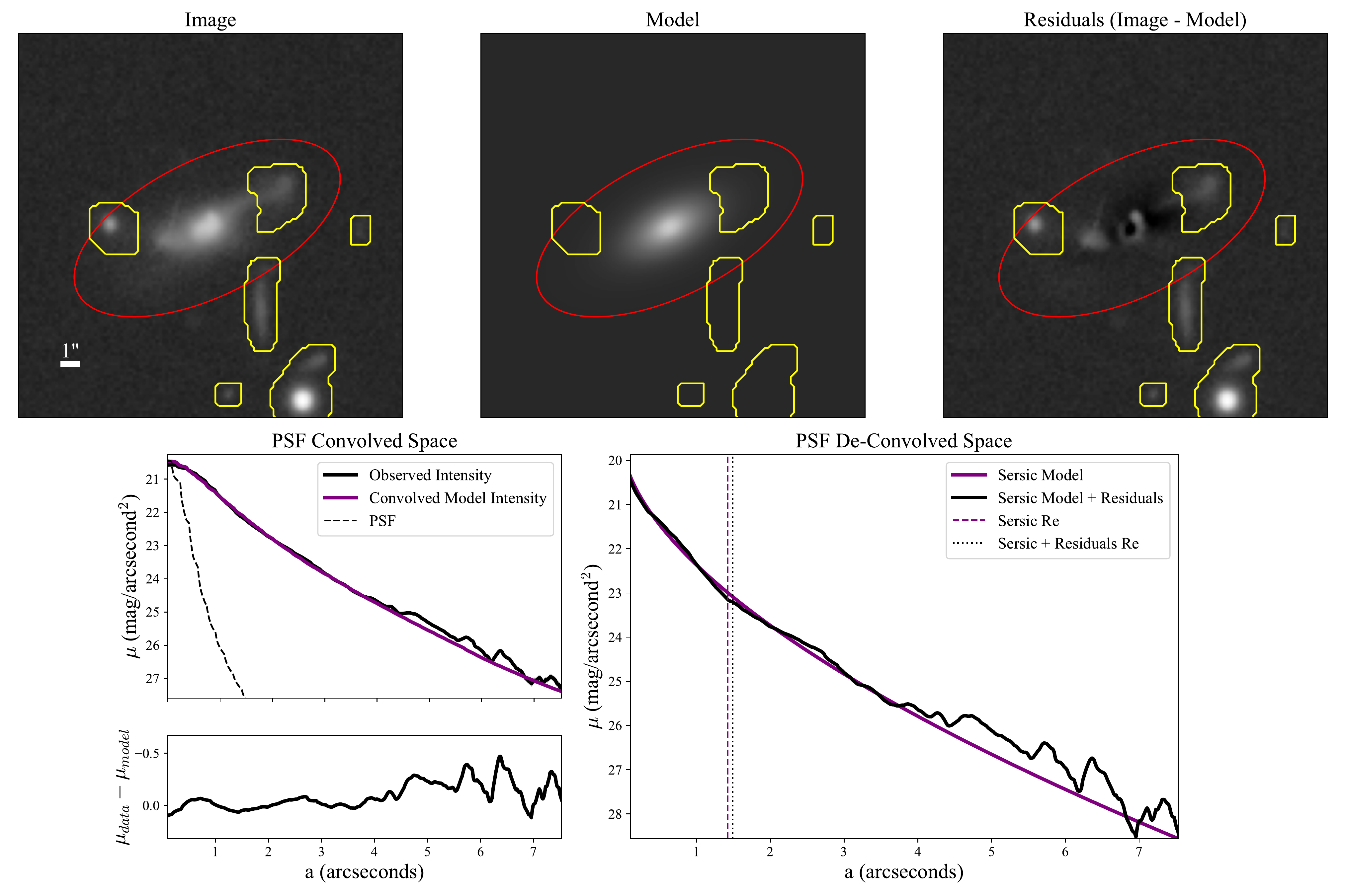}
\caption{(Top): The image, bests fitting \sersic model, and resulting residuals of a \squiggle post-starburst galaxy which exhibits clear merger features, some of which are masked (yellow outlines) in our source identification algorithms. The red aperture indicates the largest annulus used in extracting a 1D surface brightness profile, corresponding to the point where the signal-to-noise drops below 5. (Bottom Left): Observed (black) and model (purple) radially averaged surface brightness profiles as a function of the semi-major axis of the best fitting ellipse, with residuals shown below. The \sersic model under-estimates the surface brightness at large radius. (Bottom Right): The same model (purple) as the center, now shown without PSF convolution. The same profile is shown with the residuals added to them using the method from \cite{Szomoru2012}. The dashed and dotted lines represent the effective radius of the profile before and after this addition takes place. While the residuals do affect the total shape of the profile, the measurement of effective radius is largely robust to this correction. 
\label{fig:szomoru}}
\end{figure*}

We use the \texttt{source\_properties} function to extract approximate centroids, axis ratios, position angles, and fluxes of each source. We then mask all sources greater than 25 pixels (4.2 arcseconds) from the center of the image and any objects within the central 25 pixels that are more than three magnitudes fainter than the central object. We smooth the resultant mask with a 3-pixel radius top hat filter in order to ensure that we are accounting for all galaxy light from all interloping sources. These aggressive masking choices were made to ensure that all bright nearby sources are being accounted for with their own models and that the \sersic profiles are sensitive to the smooth central profiles of the galaxies. To this end, we elect in our deblending parameters to err on the side of classifying spatially distinct tidal features, such as the one to the northwest of the galaxy in Figure \ref{fig:szomoru}, as their own objects and to mask them, biasing the sizes of all galaxies we fit towards smaller values.

We then allow \texttt{GALFIT} to fit the object of interest and the remaining neighboring galaxies (those within 25 pixels and within 3 magnitudes of the central object) with \sersic profiles convolved with the PSF. We adopt the bounds on the \sersic index $0.5<\mathrm{n}<6$ following the most recent LEGA-C data release to facilitate comparisons to their fits using HST/ACS F814 images \citep{VanDerWel2021}. The centroid, position angle, axis ratio ($b/a$), and magnitudes measured by \texttt{photutils} are used as initial guesses for the parameters in the \sersic fits for each object. The sky background is initialized at 0 and is a free parameter in the fit. In Figure \ref{fig:sersic_demonstration}, we demonstrate the full \sersic fitting procedure on a \squiggle galaxy which has a nearby source that meets our criteria for simultaneous fitting. The nearby source immediately to the south of the galaxy of interest is too faint to be fit, and as such appears in the mask, but the source to the northeast is bright enough to be fit with its own \sersic model. The best fitting \sersic parameters from these fits are detailed in Table \ref{table:basic_props}.

In \cite{Greene2020}, we find that $\sim5\%$ of \squiggle post-starburst galaxies host AGN, which are identified via their strong [OIII]/$H_\beta$ ratios. However, in all cases the narrow lines indicate that the AGN are strongly obscured, and should not significantly impact the rest-frame optical continuum of the galaxies. As such, we do not elect to fit any additional point-source components to the 6 AGN in our sample. Throughout this work, we report the semi-major effective radius (\recomma) as the measure of galaxy size. 

\begin{table*}
\begin{adjustwidth}{-.6in}{-.6in}

\centering
\caption{Structural and selected spectrophotometric properties of \squiggle post-starburst galaxies}
\label{table:basic_props}
\scriptsize
\begin{tabular}{ccccccccccccc}
\hline \hline
Name & $z_{\mathrm{spec}}$ & log($\frac{M_\star}{M_\odot}$)\tablenotemark{a} & $\mathrm{r_{e,Sersic}}$ (kpc)\tablenotemark{b} & $\mathrm{r_{e,corr}}$ (kpc)\tablenotemark{c} & n\tablenotemark{b} & mag\tablenotemark{b} & b/a\tablenotemark{b} & log(\sigone)\tablenotemark{b} & G\tablenotemark{d} & GC\tablenotemark{d} & $t_q$ (Gyr) \tablenotemark{e} & Burst Fraction \tablenotemark{e}  \\ \hline

J1042+0500 & 0.6266 & 10.89 $\pm^{0.01}_{0.02}$ & 0.78 & 0.76 & 5.98 & 19.61 & 0.94 & 10.16 & 0.59 & 2.5 & 0.31 $\pm^{0.09}_{0.1}$ & 0.52 $\pm^{0.21}_{0.12}$ \\
J0907+0423 & 0.6635 & 11.27 $\pm^{0.01}_{0.02}$ & 7.03 & 6.06 & 3.47 & 19.25 & 0.85 & 9.81 & 0.5 & 2.97 & 0.11 $\pm^{0.08}_{0.04}$ & 0.36 $\pm^{0.58}_{0.2}$ \\
J0226+0018 & 0.5405 & 10.75 $\pm^{0.03}_{0.03}$ & 2.09 & 2.09 & 6.0 & 19.26 & 0.89 & 9.83 & 0.63 & 2.67 & 0.36 $\pm^{0.13}_{0.13}$ & 0.44 $\pm^{0.34}_{0.17}$ \\
J0224-0105 & 0.5979 & 10.98 $\pm^{0.01}_{0.02}$ & 3.36 & 3.21 & 5.76 & 18.82 & 0.81 & 9.97 & 0.64 & 2.84 & 0.39 $\pm^{0.18}_{0.11}$ & 0.88 $\pm^{0.12}_{0.62}$ \\
J0221-0646 & 0.6613 & 10.98 $\pm^{0.02}_{0.02}$ & 5.58 & 5.82 & 5.56 & 19.46 & 0.67 & 9.88 & 0.51 & 2.64 & 0.22 $\pm^{0.08}_{0.07}$ & 0.27 $\pm^{0.28}_{0.08}$ \\
.. & .. & .. & .. & .. & .. & .. & .. & .. & .. & .. & .. & ..
\\ \hline
\end{tabular}
\begin{tablenotes}

\tablenotetext{}{\scriptsize $^{a}$Stellar masses were derived using a delayed-exponential star formation history, described in Section \ref{subsec:squiggle}.}
\vspace{-5pt}
\tablenotetext{}{\scriptsize $^{b}$\sersic parameters were derived using single-component 2D \sersic models to fit the galaxies of interest. Uncertainties derived by refitting the galaxies using a range of PSF models from nearby locations on the sky are shown in Table \ref{table:psf_uncertainty}. For details, see Section \ref{subsec:sersic}.}
\vspace{-5pt}
\tablenotetext{}{\scriptsize $^{c}$The residual-corrected effective radii were derived using the method of \cite{Szomoru2012}, described in Section \ref{subsec:nonsersic}.}
\vspace{-5pt}
\tablenotetext{}{\scriptsize $^{d}$The Gini coefficient (G) and the generalized concentration (GC) were measured on the segmentation maps described in Section \ref{subsec:sersic}, for details see Section \ref{subsec:nonpar}.}
\vspace{-5pt}
\tablenotetext{}{\scriptsize $^{e}$Time since quenching ($t_q$) and the burst fraction are model parameters derived from non-parametric star formation history fitting of the galaxies, see \cite{Suess2022} for details}
\vspace{5pt}
\item (This table will be available in its entirety in a machine-readable form in the online journal. A portion is shown here for guidance regarding its form and content.)
\end{tablenotes}
\end{adjustwidth}
\end{table*}

Although the LEGA-C control sample is covered by the deeper HSC-Deep imaging, \squiggle galaxies fall entirely within the shallower HSC-Wide footprint. In order to compare morphological measurements fairly to \squigglecomma, we utilize imaging for the LEGA-C control sample (galaxy postage stamps and PSFs) of the HSC-Deep field that is only stacked to Wide depths. To assess the reliability of our structural measures, we compare the results from HSC fits for the LEGA-C sample to structural parameters measured from HST/ACS F814W imaging in \cite{VanDerWel2021} in Appendix \ref{sec:hst_comp}. We find that our fits are very reliable despite the images being lower in resolution by a factor of $\sim3$ compared to the HST/ACS F814W imaging. In many cases, the increased low-surface brightness sensitivity of HSC better captures the wings of galaxy profiles at large radius. This difference in profiles can explain the systematic offsets in the \sersic index and the effective radius. We find that we recover the axis ratio with extremely high precision and accuracy (median $\Delta b/a=$0.005, $\sigma_{\Delta b/a}=0.11$). 

We also test the robustness of our fits by comparing to analysis of HSC Deep imaging for the LEGA-C sample. We find that our measurements of the effective radius between the Deep and Wide are extremely consistent, with a median offset of 0\% and a scatter of $\sim8\%$ in measured size. Axis ratios are similarly well constrained, with a scatter in $\sigma_{b/a}\sim0.02$. The median \sersic index fit with the Deep images is found to be slightly lower than with Wide, with $\Delta n=-0.05$, $\Delta n_{16\%}=-0.45$, and $\Delta n_{84\%}=0.15$. We conclude that the HSC Wide images are sufficient for robustly measuring sizes of massive galaxies at intermediate redshift.

We also use the derived \sersic profiles to compute the stellar mass surface density within one kiloparsec (\sigonecomma). We do so by multiplying the fraction of the flux contained within 1 kpc by the galaxy stellar mass and dividing by the area of the bounding ellipse using the best fitting \sersic profile parameters, implicitly assuming a flat M/L gradient. 

\begin{equation}
    \mathrm{\Sigma_{1 kpc}} = \frac{M_\star \frac{f_{1 \mathrm{kpc}}}{f_{\mathrm{tot}}}}{\pi (1 \ \mathrm{ kpc})^2 (b/a)}
\end{equation}

\subsection{Systematic Size Errors Due to PSF Models} \label{subsec:psf}

The formal uncertainties on the sizes for bright galaxies measured using \texttt{GALFIT} are very small because the HSC images are so high in signal-to-noise. However, these formal uncertainties ignore potentially significant systematic errors. Likely the largest systematic is the assumption that the PSF model used in the \sersic model fitting is well determined at any point in the sky. As galaxies at intermediate-redshift have intrinsic sizes similar to the full-width-half-maximum of the point spread function, small changes in the shape of the PSF can lead to significantly different measurements of galaxy size and \sersic index.

In order to account for this, for each galaxy in \squiggle we sample the PSF from the HSC PSF Picker at 50 locations drawn from a normal distribution surrounding the galaxy with $\sigma=0.375$ degrees in RA and Dec. This choice of $\sigma$ ensures that we will be picking PSFs which are distributed throughout the entire 1.5 degree field of view of HSC surrounding the object of interest. At each position, the PSF is estimated based on surrounding stars, and the distribution in the shapes of these PSFs should approximate the uncertainty in the determination of the PSF at the position of the galaxy itself.

We refit each galaxy with each of the 50 randomly drawn PSFs and quantify the error in the measurement of the effective radius. We find that the uncertainties from this PSF shuffling are an order of magnitude larger than the formal uncertainties (median $\sigma_{\mathrm{re,formal}}=0.004$", median $\sigma_{\mathrm{re,PSF}}=0.049$"). We take these systematic errors to be dominant and throughout the rest of the work, all errors reported in the size result from this methodology. For each of the 50 iterations, we also calculate log(\sigonecomma), for which we find a median error of 0.05 dex. The full $1\sigma$ confidence intervals in \sersic properties inferred from the PSF refitting are shown in Table \ref{table:psf_uncertainty}.

\begin{figure*}
\includegraphics[width=0.48\textwidth]{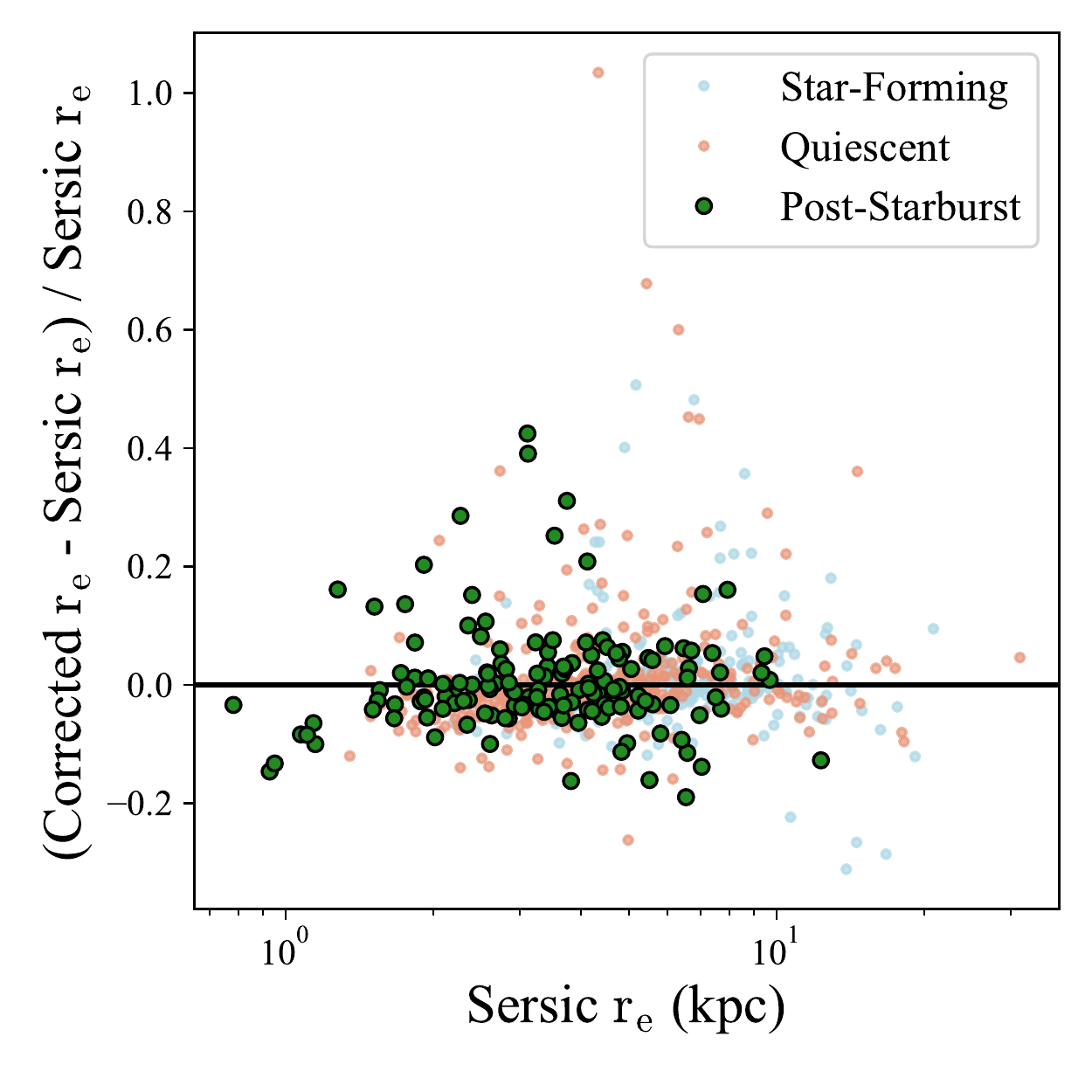}
\includegraphics[width=0.48\textwidth]{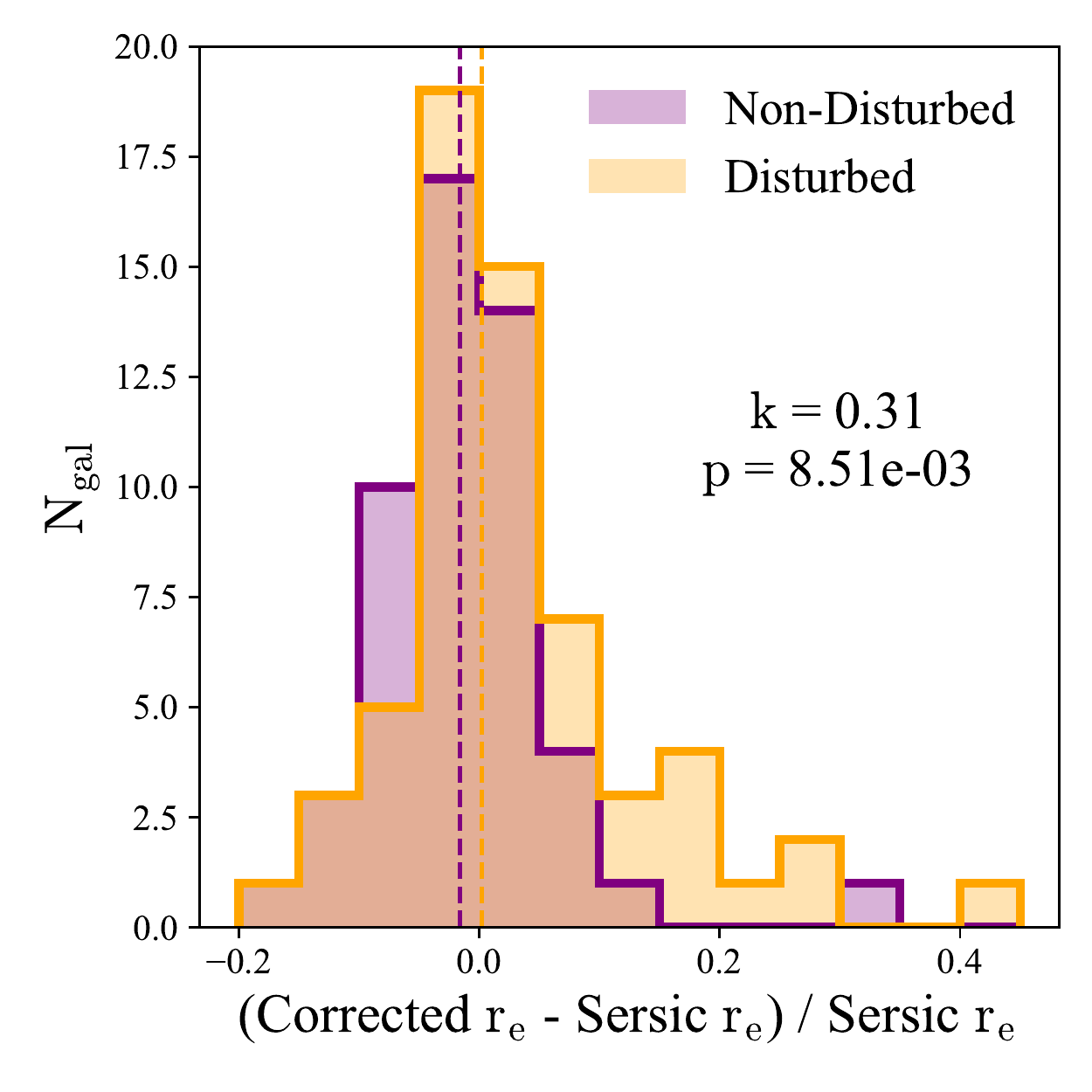}
\caption{(Left): The measured \sersic effective semi-major axis radius from \texttt{GALFIT} versus the residual-corrected percent error in \re from the application of the \cite{Szomoru2012} technique to post-starburst galaxies in \squiggle and quiescent/star-forming galaxies from the LEGA-C survey. Globally, the addition of the residuals only biases measurements to be $\sim1\%$ smaller, and majority of galaxies have measured sizes consistent within $\sim10\%$. (Right): The distributions of the residual-corrected percent error in \re for the disturbed and non-disturbed post-starburst samples. The median offsets in measured sizes (vertical lines) for the samples are comparable, but the majority of the galaxies where the corrected sizes are significantly larger are visually flagged as tidally disturbed.
\label{fig:szomoru_effectiveness}}
\end{figure*}

\subsection{Accounting for non-\sersic light} \label{subsec:nonsersic}

\sersic profiles can provide average properties of the smooth 2D light distributions of galaxies. \citep[e.g.,][]{VanDerWel2012,Almaini2017,Wu2018,Mowla2019,Kawinwanichakij2021}. However, the \sersic model does not fully account for asymmetric light from tidal features, which are present in many galaxies and are very common in the \squiggle sample of post-starburst galaxies in deep imaging (see M. Verrico et al. in preparation) and are not fully masked even in our conservative source identification procedure (see Figure \ref{fig:szomoru}). Additionally, single component \sersic fits need not perfectly describe the profiles of isolated galaxies; for example, the presence of a bright point source component (which we do not expect from AGN but could potentially result from nuclear star-formation) could drive a fit to be overly peaked, missing light in the wings of the galaxy as a result. These deviations in the \sersic fits from the true surface brightness profiles of galaxies can bias inferences about the sizes of galaxies; accounting for it is important in accurately comparing to the sizes of coeval galaxies. 

We follow \cite{Szomoru2012} to correct for deviations from \sersic profiles, including asymmetries, while accounting for PSF smearing. To start, we run \texttt{GALFIT} on galaxies as described in Section \ref{subsec:sersic}. We begin by performing annular photometry on the residual (galaxy image minus the best fitting \sersic models of \textit{all} simultaneously fit galaxies multiplied a mask which masks all non-primary sources, including those which were fit simultaneously) using the best fitting position, position angle, and axis ratios derived from the \sersic fits. We extract the profiles out to the annulus where the signal-to-noise in the galaxy image drops below 5, $\sim28$ mag/$\mathrm{arcsecond^2}$ \citep[see also][]{Huang2018}. We add these PSF-convolved residuals to the \textit{deconvolved} best fitting \sersic profile and re-measure the half-light radius from the residual corrected growth curve. An illustration of this method is shown in Figure \ref{fig:szomoru} for a clearly disturbed galaxy. Even in this extreme case, the tidal features do not contribute significantly to the total galaxy flux and the measured size is only 5\% larger than the \sersic effective radius.

In general, the residual corrected sizes do not deviate significantly from the corresponding \sersic effective radii. We show this in Figure \ref{fig:szomoru_effectiveness}a, comparing the sizes measured with \texttt{GALFIT} to the residual-corrected sizes. While we find that in a few cases, the residual-corrected sizes are significantly ($\gtrapprox25\%$) larger than those measured by \sersic fits, the median deviation from \sersic is $\sim-1\%$ for star-forming, quiescent, and post-starburst galaxies, indicating that non-\sersic and asymmetric light is not significantly affecting galaxy sizes. The majority of the measurements which skew toward significantly larger sizes are due to the presence of tidal features along the semi-major axis (see Figure \ref{fig:szomoru_effectiveness}b). We adopt these corrected effective radii as our measure of galaxy sizes for the remainder of this work. However, due to the lack of systematic offsets in any of the samples, all conclusions in this work would not change if we were to use the \sersiccomma-only half-light sizes. 

\section{The Sizes and Structures of Post-Starburst Galaxies} \label{sec:structure}

\begin{table*}

\centering
\caption{Uncertainties on \sersic parameters from PSF model refitting}
\label{table:psf_uncertainty}
\scriptsize
\begin{tabular}{cccccccccccccccc}
\hline \hline
Name & & $\mathrm{r_e}$ (kpc) & & & n & & & mag & & & b/a & & & $\log(\Sigma_{1\mathrm{kpc}})$ & \\
& 16\% & 50\% & 84\% & 16\% & 50\% & 84\% & 16\% & 50\% & 84\% & 16\% & 50\% & 84\% & 16\% & 50\% & 84\% \\ \hline

J1042+0500 & 0.94 & 1.3 & 1.62 & 2.72 & 3.09 & 5.82 & 19.64 & 19.66 & 19.67 & 0.74 & 0.82 & 0.94 & 9.94 & 10.02 & 10.1 \\
J0907+0423 & 6.98 & 7.3 & 8.28 & 3.21 & 3.51 & 4.64 & 19.16 & 19.24 & 19.26 & 0.84 & 0.86 & 0.89 & 9.71 & 9.73 & 9.8 \\
J0226+0018 & 2.25 & 2.37 & 2.48 & 4.02 & 5.32 & 6.0 & 19.25 & 19.27 & 19.31 & 0.82 & 0.86 & 0.9 & 9.69 & 9.74 & 9.76 \\
J0224-0105 & 3.04 & 3.22 & 3.29 & 4.86 & 5.63 & 6.0 & 18.82 & 18.84 & 18.87 & 0.78 & 0.8 & 0.83 & 9.86 & 9.88 & 9.91 \\
J0221-0646 & 5.39 & 5.58 & 5.83 & 5.16 & 5.6 & 6.0 & 19.44 & 19.45 & 19.47 & 0.68 & 0.71 & 0.74 & 9.68 & 9.7 & 9.72 \\
.. & .. & .. & .. & .. & .. & .. & .. & .. & .. & .. & .. & .. & .. & .. & ..
\\ \hline
\end{tabular}
\begin{tablenotes}
\item (This table will be available in its entirety in a machine-readable form in the online journal. A portion is shown here for guidance regarding its form and content.)
\end{tablenotes}
\end{table*}

\begin{figure}
\includegraphics[width=0.45\textwidth]{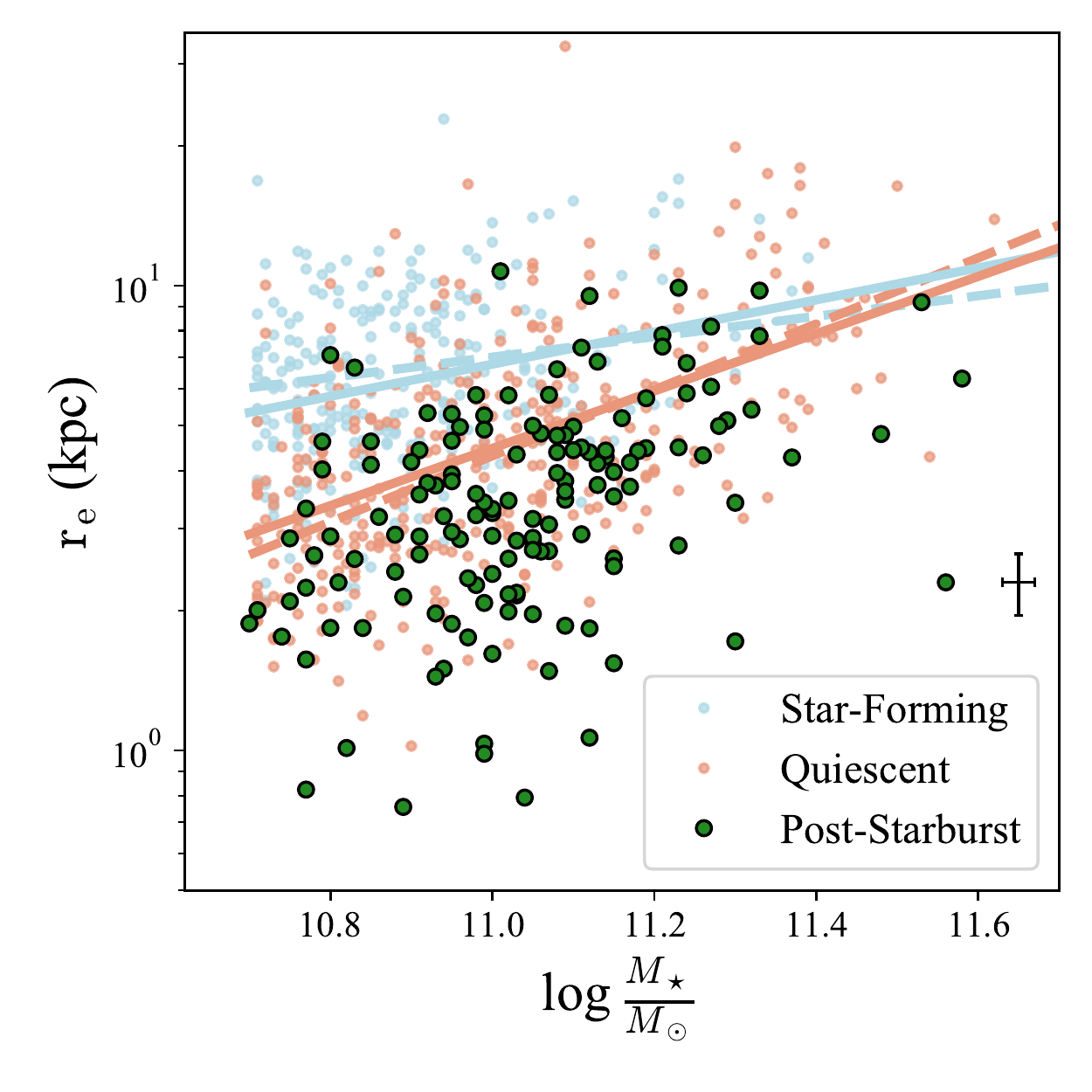}
\caption{The effective radius versus stellar mass relation for $z\sim0.7$ post starburst (green), quiescent (red), and star-forming (blue) galaxies. Characteristic error bars in the mass and size are shown in black. The best fitting relations to star-forming and quiescent galaxies (blue and red respectively) are shown as solid lines. Best fits from \cite{VanDerWel2014} for $0.5<z<1.0$ are shown as dashed lines. \squiggle post-starburst galaxies are compact on average relative to both star forming and quiescent galaxies, though there is significant scatter within each population.
\label{fig:mass_size}}
\end{figure}

\begin{figure*}
\includegraphics[width=0.33\textwidth]{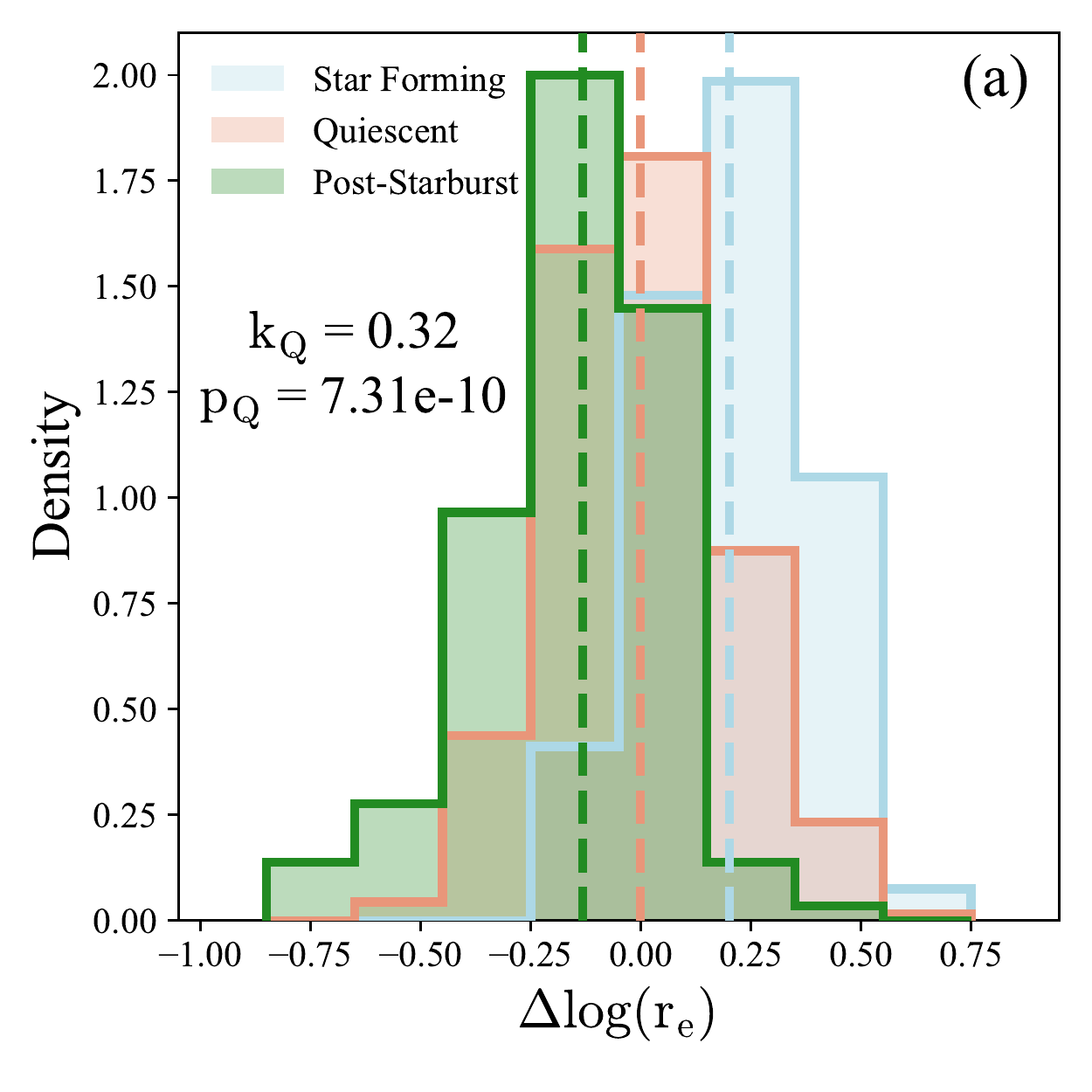}
\includegraphics[width=0.33\textwidth]{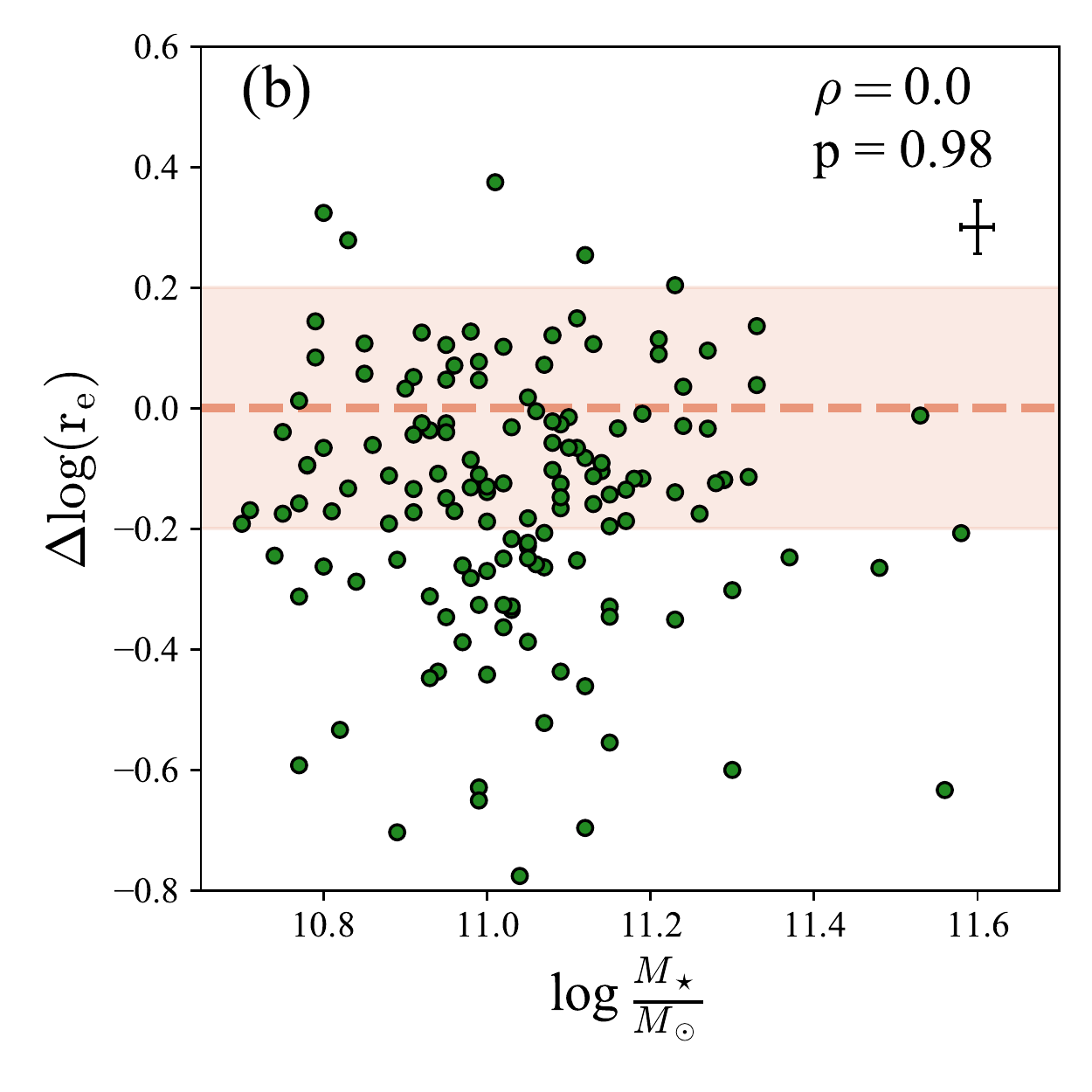}
\includegraphics[width=0.33\textwidth]{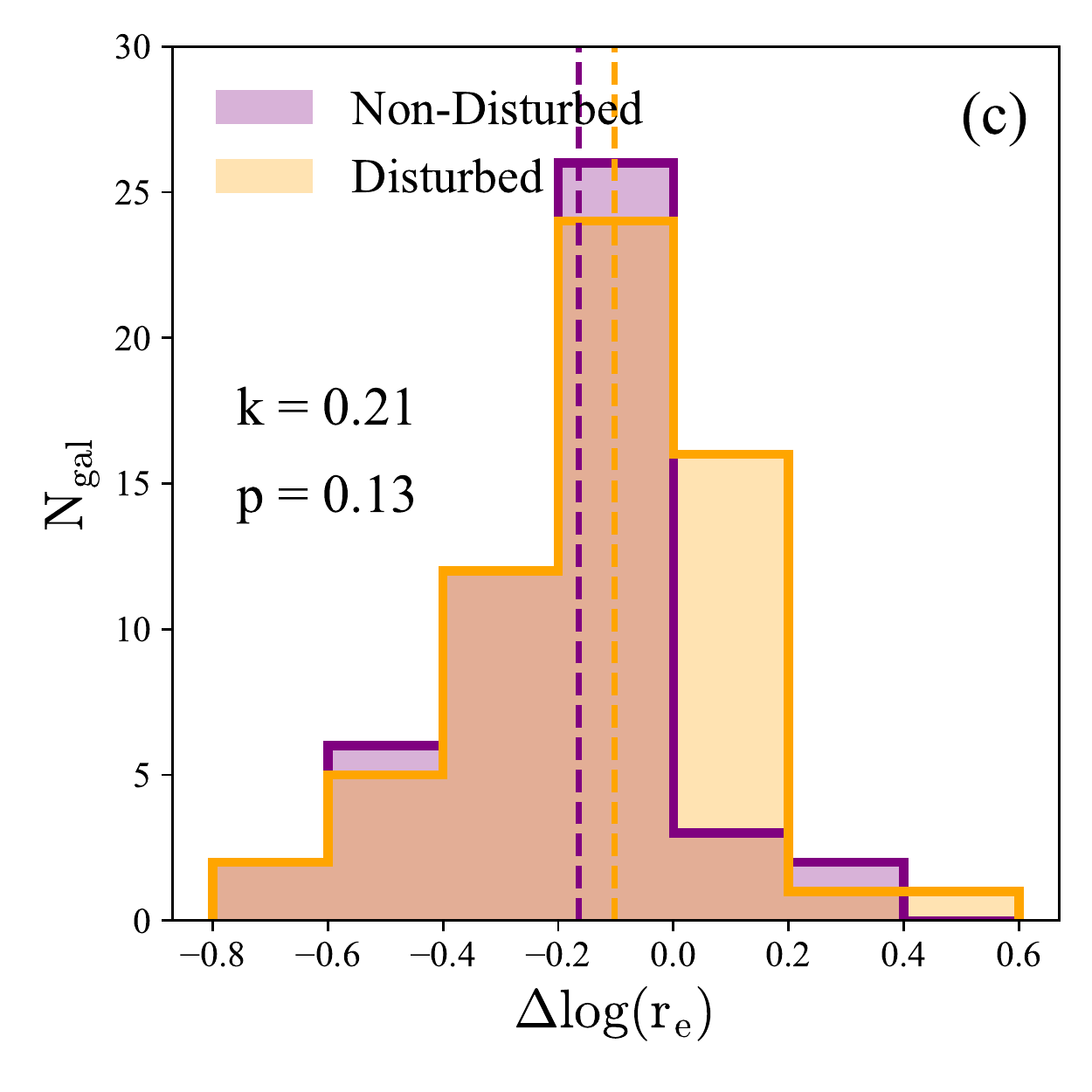}
\caption{(Left): The distribution of \deltare from the quiescent mass-size relation for the star-forming (blue), quiescent (red), and post-starburst (green) samples. The medians of the samples are indicated with vertical dashed lines. The star-forming sample is a median $\sim$ 0.2 dex larger than the quiescent sample at a given stellar mass. The post-starburst sample is a median of 0.13 dex below the quiescent population. All 3 samples have similar scatter of $\sim$ 0.2 dex. (Center): \deltare versus log $M_\star$ for the \squiggle post-starburst sample (green). The red dashed line and shaded region represent the quiescent mass-size relation. Typical errors are shown as in black. Post-starburst galaxies uniformly scatter below quiescent galaxies, with no trend as a function of the stellar mass as evidenced by the Spearman correlation coefficient ($\rho$) and associated p-value. (Right): Distributions in \deltare for the non-disturbed an disturbed samples of post-starburst galaxies as defined in Verrico et al. in preparation. While the median size at fixed stellar mass is slightly smaller than for the non-disturbed sample, a Kolmogorov–Smirnov test shows that the samples are consistent with being drawn from the same distribution (see associated k and p values). 
\label{fig:delta_re_hists}}
\end{figure*}

\subsection{Post-starburst half-light sizes}

Across redshift, star-forming and quiescent galaxies exhibit differing relations in the size-mass plane when size is quantified by the half-light radius \citep{VanDerWel2014,Wu2018,Mowla2019}. The sizes of galaxies scale with their stellar mass, but quiescent galaxies are generally smaller and follow a steeper relation than star-forming galaxies. In this section, we hope to constrain the rapid path to quiescence by comparing the sizes of coeval post-starburst, quiescent, and star-forming galaxies. In Figure \ref{fig:mass_size}, we show the residual corrected effective radius versus mass for the \squiggle post-starburst galaxies in green, along with the mass-representative ($\mathrm{log}\frac{M_\star}{M_\odot}>10.7$) LEGA-C star-forming (blue) and quiescent (red) samples. The blue and red solid lines are our best fitting relations for star-forming and quiescent galaxies respectively, and the dashed lines are the best fitting relations from \cite{VanDerWel2014}. There is considerable scatter in the sizes of the post-starburst galaxies, but on average they are fairly compact relative to all of the LEGA-C sample.

In order to test the relative compactness of post-starburst galaxies, we define \deltarecomma, the vertical offset with regard to the quiescent mass-size relation, $\mathrm{\hat r_{e, Q}}(\log \frac{M_\star}{M_\odot})$, as follows:

\begin{equation}
    \mathrm{\Delta \log(r_e)} = \log(\mathrm{r_e}) - \log(\mathrm{\hat r_{e, Q}}(\log \frac{M_\star}{M_\odot}))
\end{equation}

In Figure \ref{fig:delta_re_hists}a, we show the distributions in \deltare of the \squiggle post-starburst galaxies in green in addition to the quiescent (red) and star-forming (blue) mass-complete samples. All three populations show similar scatter ($\sim0.2$ dex), but the \squiggle galaxies $\sim$0.1 dex more compact in their light distributions than quiescent galaxies at similar redshift. In order to quantify the uncertainty in the inferred median \deltarecomma, we refit the quiescent mass-size relation using the \texttt{emcee} implementation of Markov Chain Monte Carlo \citep{Foreman-Mackey2013} and draw randomly from the posterior for the intercept and slope. We find a $1\sigma$ confidence interval on the median value of \deltare of [-0.17, -0.07]. In Figure \ref{fig:delta_re_hists}b, we show that the offset from this sequence does not correlate with the mass of the galaxy. We perform a two-sample Kolmogorov–Smirnov test on \deltare for the post-starburst and quiescent samples find that they are not consistent with being drawn from the same distribution (p=2.57e-4). In addition, we perform a 2D Kolmogorov–Smirnov test \citep{Peacock1983, Fasano1987} on the mass-matched mass-size plane to ensure that this conclusion is robust to our definition of \deltarecomma. This test confirms that post-starburst galaxies are not consistent with being drawn from the same 2D distribution as the quiescent galaxies (p=7.95e-4). 

The compact sizes we measure for the post-starburst sample relative to coeval quiescent and star-forming populations are broadly consistent with the smaller-than-average coeval post-starburst galaxies selected from the LEGA-C survey \citep{Wu2018, Wu2020} in addition to high mass ($M_\star>10^{10} \ M_\odot$) post-starburst galaxies at $1<z<2$ \citep{Yano2016,Almaini2017, Maltby2018}. The light-weighted sizes are also similarly compact to the $M_\star\sim10^{11} \ M_\odot$ post-starburst galaxies in \cite{Suess2020}. In Figure \ref{fig:delta_re_hists}c, we show the distributions in \deltare splitting the sample into disturbed and non-disturbed using the classifications from Verrico et al. in preparation. While the median size at fixed stellar mass is slightly larger for the disturbed population (owing to the fact that tidal features do influence the effective radius), a Kolmogorov–Smirnov test shows that the distributions do not differ significantly, and the entire post-starburst sample is compact relative to coeval quiescent galaxies. 

Crucially, we note that these galaxies are resolved in the HSC imaging. In Appendix \ref{sec:hst_comp}, we demonstrate that the sizes we measure using HSC imaging of galaxies at this redshift are almost entirely consistent with the sizes measured on the smallest galaxies in \cite{VanDerWel2021}, and in Appendix \ref{sec:resolved_check}, we demonstrate that the 1D and 2D surface brightness profiles of the smallest galaxies in \squiggle are significantly different from that of the PSF. In the i-band, the light of the youngest stellar population will almost completely dominate over any older stellar population, and so the sizes we measure are likely to primarily trace the physical extent of that population. This finding indicates that the recent star formation was extended on at least kpc scales past the circumnuclear region in all the galaxies in our sample.

\begin{figure*}
\centering
\includegraphics[width=0.85\textwidth]{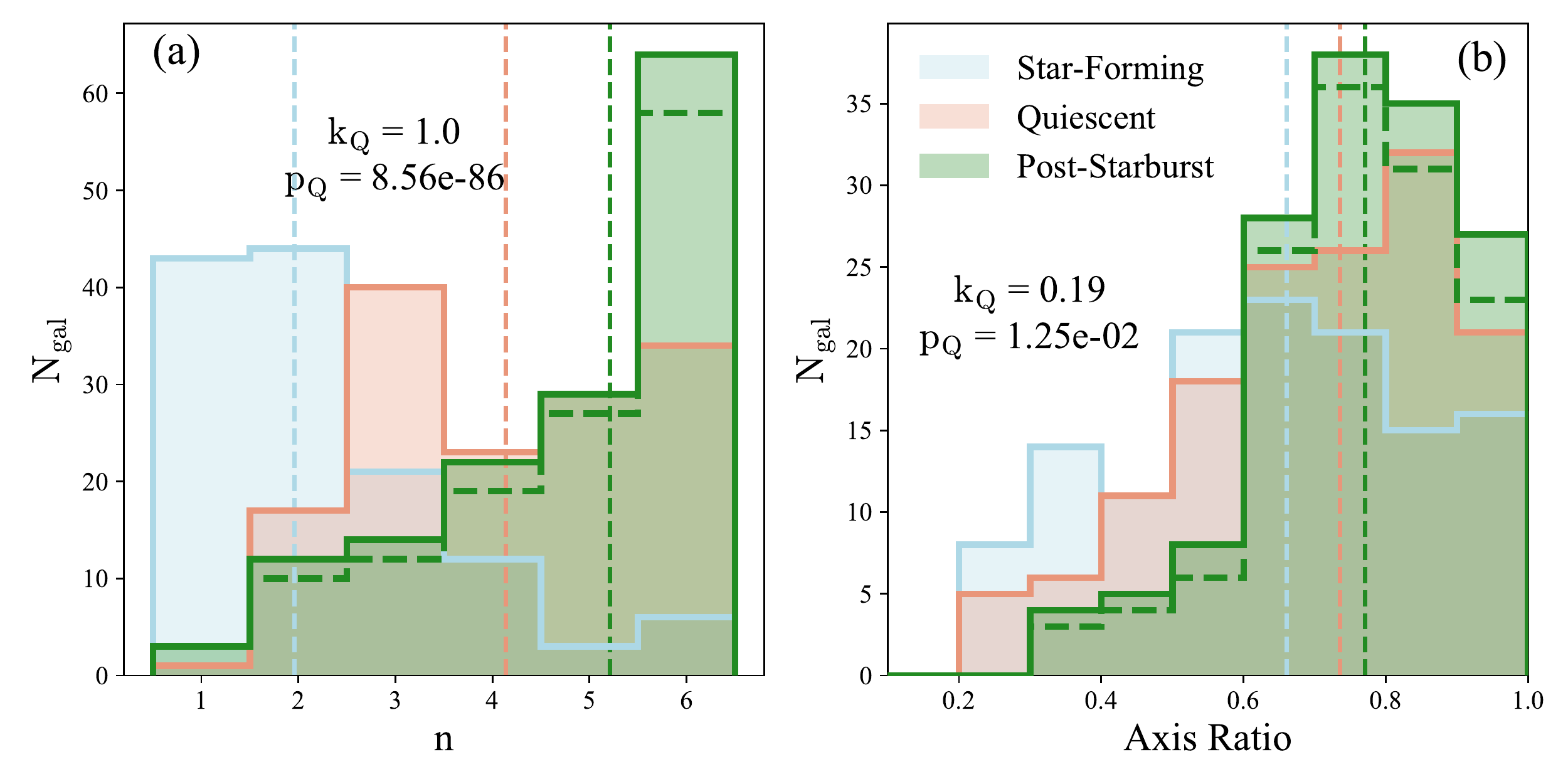}
\caption{(Left) The distributions of mass-matched post-starburst (green), star-forming (blue), and quiescent (red) galaxies for the \sersic index $n$. The dashed post-starburst histogram shows the post-starburst galaxies which are mass-matched to the star-forming sample. Post-starburst galaxies are overwhelming fit with large n, often running up against the $n=6$ boundary. The \sersic index distribution is distinct from that of star-forming galaxies, which favor small n. (Right) The distributions of axis ratios, defined as the ratio of the semi-minor axis to the semi-major axis. Star-forming galaxies show the most elongation, followed by quiescent galaxies, and post-starburst galaxies are slightly rounder than both. In both plots, the median of each distribution is shown as a dashed vertical line. The results of a Kolmogorov–Smirnov test between the quiescent and post-starburst distributions are shown on both figures.
\label{fig:n_ar_hists}}
\end{figure*}

\subsection{Other parametric measures of structure} \label{subsec:par}

In addition to the half-light radius, the \sersic models also include the \sersic index (n) and the projected axis ratio ($b/a$) in the 2D fits to the galaxies. These structural measures encode information about the 3D structures and light profiles of galaxies and show different empirical trends for star-forming and quiescent populations \citep[e.g.][]{Shen2003, VanDerWel2014}. In Figure \ref{fig:n_ar_hists}, we show the distributions in the \sersic 
index and the axis ratio for post-starburst galaxies, as well as the mass-matched quiescent and star-forming control samples. As expected, the star forming sample and the quiescent sample are significantly different in their \sersic index distributions; star forming galaxies are fit with small \sersic index ($\mathrm{n_{med}}\sim2$) and quiescent galaxies are systematically fit with higher n ($\mathrm{n_{med}}\sim4$). Post-starburst galaxies have higher \sersic indices than both of these samples, with a median \sersic index of 5.2. These higher \sersic indices are likely driven by the concentrated light of these objects, and visual inspection of the median 1D surface brightness profiles of the sample confirms that the majority of the galaxies are being well fit by these high \sersic indices. We note, however, that many post-starburst and quiescent galaxies run up against the boundary we impose at n=6. We run a number of tests including residual inspection, fitting images with a central point source component in addition to a \sersic profile, and expanding the threshold \sersic index to n=8, in order to quantify the effects of the run-up against the boundary to our results. Ultimately, the qualitative results of the paper are insensitive to any changes to our fitting procedure, but we describe these tests below.

For the sample of \squiggle post-starburst galaxies, 36\% galaxies run up against the n=6 boundary, whereas in the mass-matched quiescent sample 20\% of the galaxies are best fit with n=6. In general, the \sersic index is the least well constrained property in our fits, as it is extremely sensitive to the shape of the PSF model we use in fitting. Additionally, these uncertainties grow as a function of the best fitting \sersic index, such that for galaxies with intrinsically high \sersic indices (e.g., quiescent and post-starburst galaxies), the \sersic indices are particularly poorly constrained. Visual inspection of the residual 1D surface brightness profiles of the galaxies which run up to the n=6 boundary show that, in contrast with the rest of the sample, these have median central mismatches of $-0.04 \ \mathrm{mag/arcsecond^2}$. The models are systematically fainter in the center than the galaxies and brighter than the galaxies on scales of $\sim1$". This imperfect agreement between data and model is what we would expect from a mismatch between the model PSF we use in the fitting of the galaxies and the true PSF of the galaxies, where the centrally concentrated light of the galaxies is smeared by a PSF which is wider than the true PSF. In some cases (e.g. galaxy J0226+0018 in Table \ref{table:basic_props} and \ref{table:psf_uncertainty}), a nearby PSF model is capable of causing the fit to converge to a \sersic index which is within the range we allow, but in many cases all 50 fits in our rerun with nearby PSF models converge to the same n=6 value.

This issue of high-n pileup cannot be resolved by simply expanding the \sersic index to higher values. We re-run our fits of the \squiggle sample with an upper boundary on the \sersic index of n=8 and find that 20\% of the galaxies still converge to the upper boundary and show the same characteristic residual shape as in the n=6 boundary fits, indicating that the light is generally not well fit by a \sersic profile regardless of the limit we place on n. For the galaxies which were best fit with n=6 in the original fits, these new fits yield residual corrected sizes which are a median of 10\% larger than those fit with n=6, which is a significant change but is not one which would significantly alter any of our conclusions about the size of the sample relative to co-eval galaxies. In addition, we also run a set of fits with an additional point source component fixed to the same centroid as the galaxy of interest. $\sim25\%$ of these fits fail to converge, and among those that do converge, the majority are fit with an essentially negligible point-source component (median \sersic mag - point-source mag = -2.47). Additionally, the galaxies which are fit with significant with point source components tend not to be those which are at the n=6 boundary. We conclude that deviations from \sersic profiles are likely being driven by PSF models which are not perfectly matched to the intrinsically high \sersic index galaxies we are fitting. On the whole, all conclusions regarding the size of the galaxies are not strongly influenced by the inability to perfectly determine the \sersic index, as evidenced by the very good match in the sizes of the LEGA-C quiescent galaxies we measure to those measured using HST/ACS imaging (see Appendix \ref{sec:hst_comp}).


The distribution of projected axis ratios for \squiggle post-starburst galaxies also skews significantly higher star-forming galaxies and slightly higher than quiescent galaxies. Taken together, these parametric measures indicate that post-starburst galaxies structurally appear to be fairly compact spheroids. If these galaxies began their lives as extended disks, any structural transformation must have pre-dated or occurred concurrently with the shutdown of star formation, such that they structurally compact $\sim100$ Myr after quenching. 


\begin{figure}[]
\includegraphics[width=0.45\textwidth]{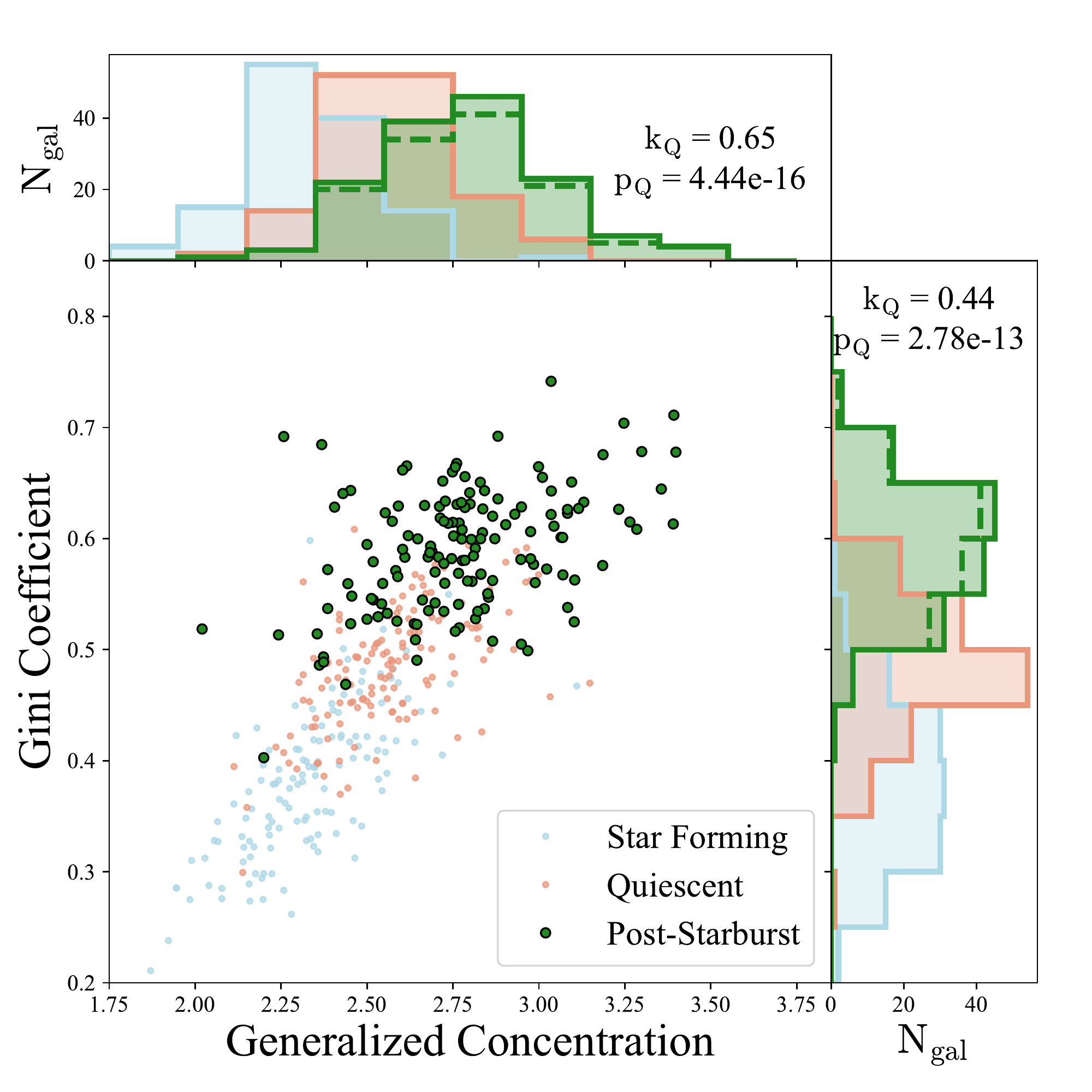}
\caption{The Gini coefficient versus the generalized concentration for the mass-matched post-starburst (green), quiescent (red), and star-forming (blue) galaxies. The dashed post-starburst histogram shows the post-starburst galaxies which are mass-matched to the star-forming sample. These non-parametric measures of light concentration are strongly correlated. By both measures, the post-starburst population is more concentrated than both the quiescent and star-forming populations, in agreement with \sersic parameters.
\label{fig:gini_gc}}
\end{figure}

\subsection{Non-parametric measures of structure} \label{subsec:nonpar}

Although we have shown that the light profiles of the post-starburst galaxies and the coeval control sample do not significantly deviate from \sersiccomma, the assumption of \sersic profiles, especially those which run up against an artificial barrier in \sersic index n, may introduce model dependent effects that muddy interpretations of compactness. Therefore, as a parallel test, we turn to non-parametric measures, namely the Gini coefficient and the generalized concentration (GC), which we measure using the \texttt{GALMORPH} suite \citep{Freeman2017}. We use the standard definition of the Gini coefficient defined in \cite{Abraham2003}. However, the definition of the generalized concentration differs from the traditional concentration statistic \citep[e.g.][]{Conselice2003} by not using circular apertures but instead by comparing the minimum number of pixels which contain 20\% of the galaxy's light ($a_{20}$) to the minimum number of pixels containing 80\% of the galaxy's light ($a_{80}$) as follows:

\begin{equation}
    \mathrm{GC} = 5 \log_{10} \frac{a_{80}}{a_{20}}
\end{equation} 

These measures rely only on the rank ordering of pixels after source identification is performed. Higher values of these parameters indicate a high concentration of light in very few pixels.

In Figure \ref{fig:gini_gc}, we show the distributions of the post-starburst and control samples in this space for the mass-matched samples. The two measures are indeed correlated, and confirm the compact nature of post-starburst galaxies relative to the quiescent and star forming control samples. The post-starburst galaxies are particularly distinct from star-forming galaxies in Gini-concentration plane, indicating that if these galaxies are the descendants of similarly extended star-forming galaxies, they must have undergone a significant structural transformation. This is the same conclusion we draw from the parametric metrics (see Section \ref{subsec:par}).

\section{The origins of compact post-starburst galaxy structure \label{sec:discussion}}

In this section, we speculate on the origin of the compact structure and identify possible progenitors to the rapid channel of quenching.

\subsection{Testing the central starburst scenario \label{subsec:centralburst}}

\begin{figure*}
\includegraphics[width=\textwidth]{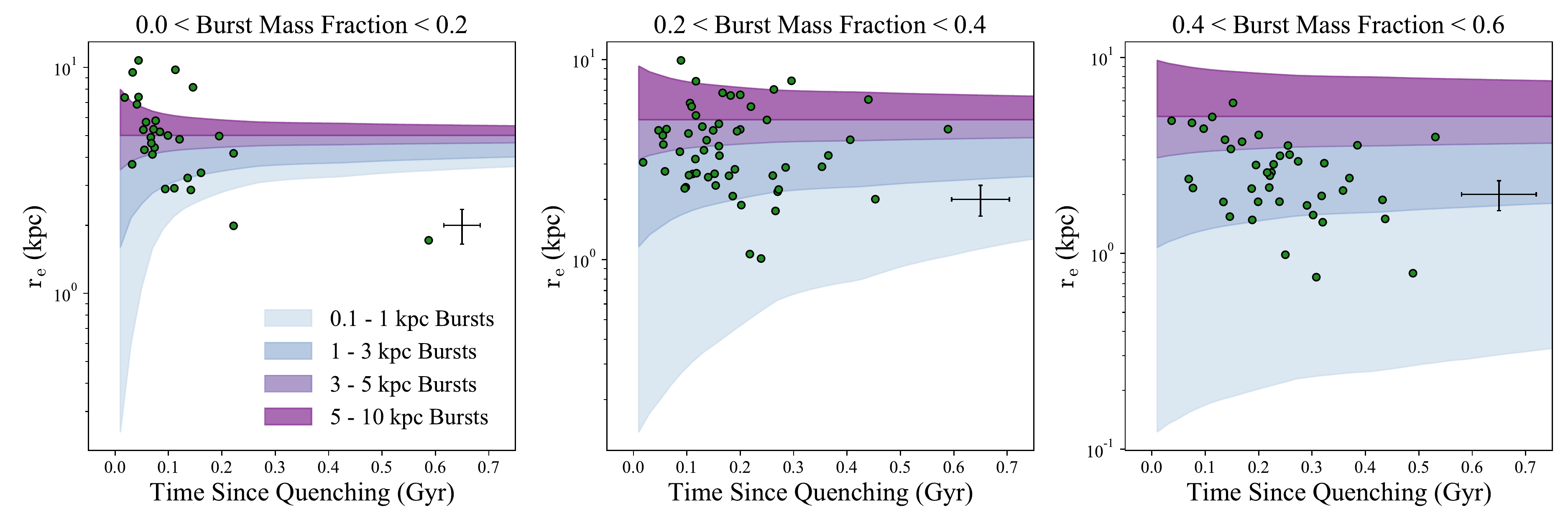}
\caption{The half-light radius versus the time since quenching for the \squiggle post-starburst galaxies, divided into three bins of burst mass fraction (as measured in \cite{Suess2022}). In the background, we shade the regions of size evolution populated by burst models described in Section \ref{subsec:centralburst} for a range of central burst sizes. We find that the majority of \squiggle post-starburst galaxies are inconsistent with the models of a sub-kpc scale burst, and can only be well-described by bursts that are extended on 1-10 kpc scales. \label{fig:re_tq}}
\end{figure*}

Quenching caused by merger-driven central starbursts resulting in quenched galaxies have been shown to occur in simulations \citep[e.g.][]{Bekki2005,Wellons2015}, and gas-rich wet compaction events could similarly cause a centrally concentrated starburst which, post-burst, would appear compact \citep{Tacchella2016, Zolotov2015}. Observations of post-starburst galaxies have suggested that their compact structures are consistent with being the result of a recent central starburst \citep{Wu2018, Wu2020, DeugenioF2020}. In addition, at redshifts similar to those of \squigglecomma, extremely compact starburst galaxies have been observed with $\sim25\%$ of the mass of $10^{11} \ M_{\odot}$ galaxies existing in the central $\sim100$ parsecs \citep{Sell2014, Diamond-Stanic2021}. 

Our previous study of 6 galaxies in the \squiggle sample using spatially resolved spectroscopy measured flat age gradients, disfavoring formation via a secondary central burst of star formation \citep{Setton2020}. However, if the recent burst of star formation totally dominates the light of the galaxy, hiding existing gradients, we would expect those signatures to fade with time. As the \squiggle sample spans a range of 0-700 Myr since quenching, we can use it to test whether galaxies are consistent with centralized starbursts imposed on older, more extended populations. Because the youngest, brightest stars in newly formed stellar populations die the most quickly, a central starburst's influence on the size of a galaxy would become weaker with time, leading to size growth as a function of the time since quenching. In Figure \ref{fig:re_tq}, we show a subset of the \squiggle sample divided up into bins of burst mass fraction (as measured in \cite{Suess2022}) to show trends for galaxies which formed a similar fraction of their stellar mass in the recent burst of star formation. None of these bins show the predicted positive trend for a central starburst; conversely, the smallest and largest burst fraction bins are consistent with a negative slope (Spearman correlations for the 0-20\% bin: $\rho=-0.52, \mathrm{p}=0.004$; Spearman correlations for the 40-60\% bin: $\rho=-0.38, \mathrm{p}=0.01$), and the central bin is consistent with no slope (Spearman correlation for the 20-40\% bin: $\rho=-0.10, \mathrm{p}=0.45$). We additionally test for correlations between the Gini coefficient and the generalized concentration parameters and find no significant correlation with the time since quenching. The lack of a positive correlation between these measures of concentration and the time since quenching suggests that highly centralized star formation is unlikely to have occurred prior to quenching in post-starburst galaxies.

We further illustrate this result via the use of a toy model of half-light radius evolution. Using \texttt{fsps} \citep{Conroy2009, Conroy2010, ForemanMackey2014} to track the evolution of the rest frame optical light with time, we simulate the superposition of an instantaneous burst of star formation on an older galaxy population by modeling both the galaxy and the burst as n = 4 \sersic profiles. We fix the older population to be 1 Gyr old and to have \re = 5 kpc, the approximate size of a $10^{11} M_\odot$ quiescent galaxy at this redshift. We test 3 burst regimes, 0 - 20\% burst fraction (modeled with a 10\% burst), 20 - 40\% burst fraction (modeled with a 30\% burst), and 40 - 60\% burst fraction (modeled with a 50\% burst), and a range of burst geometries ranging from an extremely centralized 100 parsec burst to an extended 10 kpc burst. We set the central and underlying dust attenuation to 0.5 and 0.1 dex respectively using the \texttt{fsps dust2} parameter to reflect that the central region may be significantly more attenuated than the older population, but still restricting to the range of best fitting dust values from spectrophotometric modeling (median \texttt{dust2} $\sim$ 0.3, see \citealp{Suess2022}). We note that these assumptions neglect the possibility that heavily obscured star formation is contained in the galaxies. Some post-starburst galaxies have been shown to contain deeply embedded dust reservoirs in the central $\sim100$ parsecs with $A_\mathrm{v}\sim10^4$, which could in principle shield large amounts of central star formation \citep[][]{Smercina2021}. However, in a previous study of CO(2-1) for a small (13 galaxy) subsample of \squigglecomma, we do not see evidence of centrally concentrated molecular gas or continuum emission in the detected galaxies at $\sim1"$ resolution. We measure molecular gas effective radii on the order of kiloparsecs and do not detect continuum emission in all but one galaxy \citep{Bezanson2022}. As such, we assume dust obscuration informed by our best fits to OIR spectrophotometric data.

We show the tracks generated by these models as shaded regions on Figure \ref{fig:re_tq}. The vast majority of \squiggle galaxies are inconsistent with a sub-kpc scale burst of star formation. To ensure that these conclusions are not heavily dependent on toy model assumptions about the underlying galaxy age and size, we test for a range of values (\re = 3 kpc and 10 kpc, age = 0.5 Gyr) and find that the majority of galaxies are too large to ever overlap with the sub-kpc burst tracks. We note that these observed kpc-scale bursts of star formation do not completely eliminate mergers as a possible trigger for inducing the burst and subsequent suppression of star formation, as IFU studies of post-merger galaxies have found evidence for centrally peaked but still global enhancements in star formation rate \citep{Thorp2019}. The lack of agreement between models for a highly concentrated burst of star formation and the size-age trends we observe only shows that star formation could not have occurred solely in the central regions of recently quenched galaxies.

The negative/inconclusive trends in \re versus time since quenching indicate that these post-starburst galaxies do not evolve significantly in their $\sim500$ Myr after quenching and that the youngest stars in the galaxy are distributed on spatial scales similar to any underlying older population. This is consistent with the finding that a small sample of \squiggle post-starburst galaxies exhibit flat age gradients \citep{Setton2020}, which are also seen in young quiescent galaxies closer to cosmic noon \citep{Akhshik2020,Jafariyazani2020} and compact local quiescent galaxies that may be the descendants of early universe rapid quenching \citep{Schnorr-Muller2021}. As we are still accounting for some unmasked tidal light in the fits to our galaxy sizes (see Figure \ref{fig:szomoru}), we propose that any negative trend in time since quenching versus size may be the result of tidal features which are commonly present in the young post-starburst galaxies \citep[e.g., M. Verrico et al. in preparation;][]{Sazonova2021} and which fade on $\sim200$ Myr timescales. There is evidence for this when we split the sample into disturbed and non-disturbed using the classifications from M. Verrico et al. in preparation, as the youngest galaxies in the sample are far more likely than the oldest ones to host visible tidal features. However, we note that the sizes of the galaxies we fit are not being primarily driven by asymmetric non-\sersic light, as the corrections for the disturbed and non-disturbed samples are both consistent with no systematic offset in measured half-light radius, and instead are primarily driven by the central light of the galaxy due to our aggressive deblending and masking. 

\begin{figure*}
\includegraphics[width=0.49\textwidth]{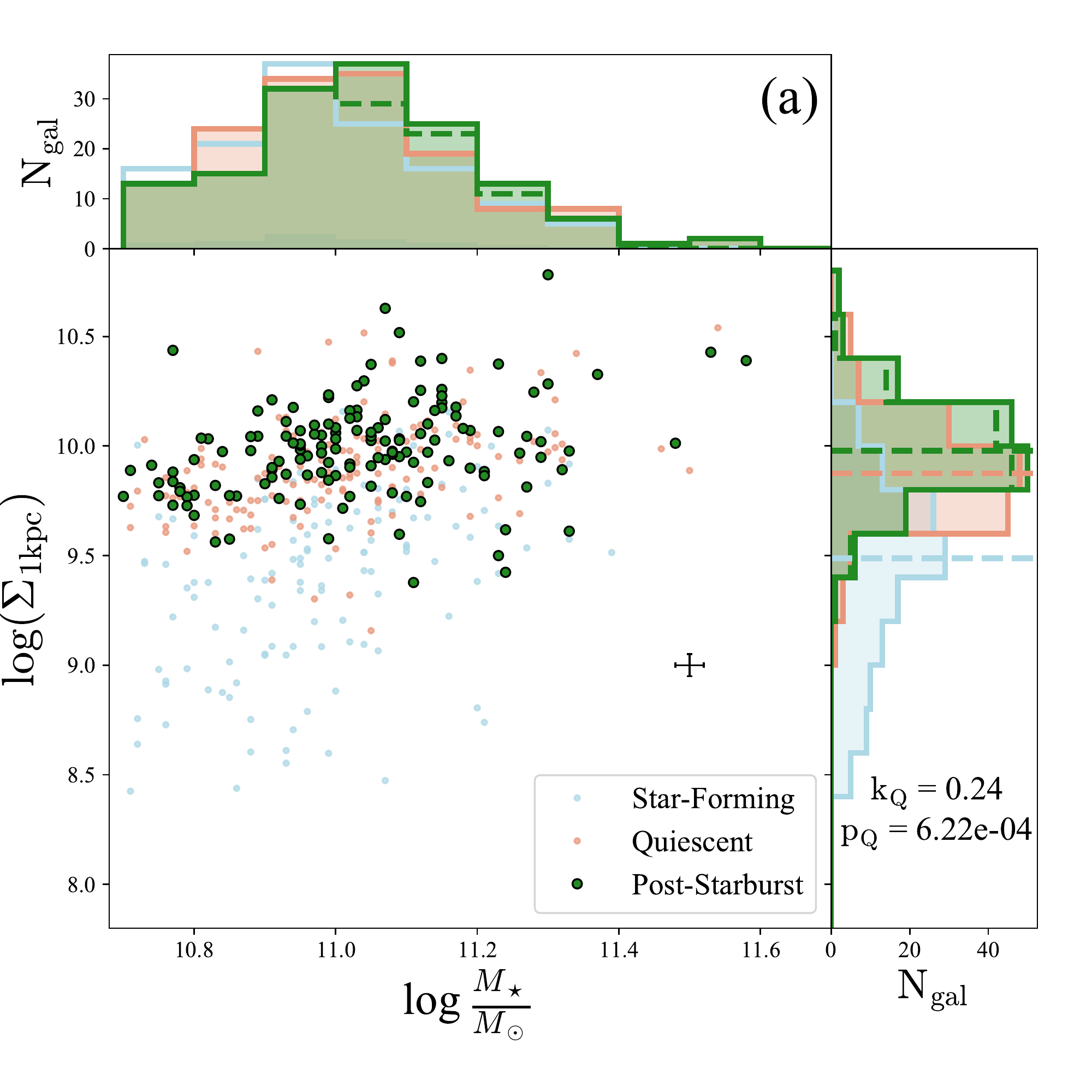}
\includegraphics[width=0.49\textwidth]{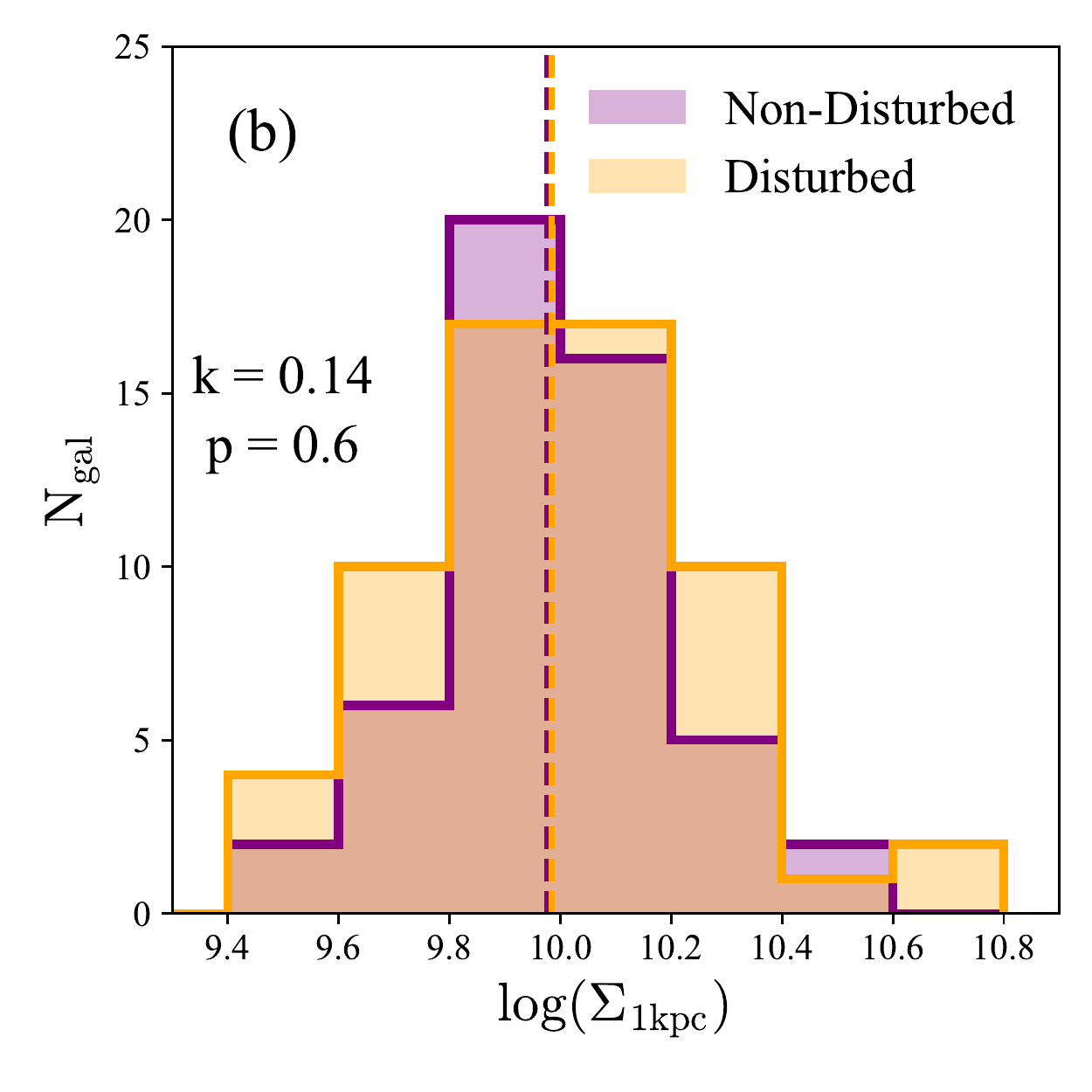}
\caption{(Left): The stellar mass surface density in the central kpc as a function of the stellar mass for \squiggle along with the star-forming and quiescent sub-samples. Typical error bars derived from the PSF shuffling procedure are shown in black. The dashed post-starburst histogram shows the post-starburst galaxies which are mass-matched to the star-forming sample. \squiggle galaxies are only slightly denser than quiescent galaxies, while star-forming galaxies are significantly less dense than both populations. The results of a Kolmogorov–Smirnov test between the quiescent and post-starburst log(\sigonecomma) distributions are shown on the verticle histogram. (Right): The distributions of in log(\sigonecomma) for the disturbed and non-disturbed post-starburst populations, as defined in Verrico et al. in preparation. As with the compactness at fixed stellar mass (see Figure \ref{fig:delta_re_hists}), a Kolmogorov–Smirnov test shows that the distributions are not significantly different between the two populations. The \squiggle sample of post-starburst galaxies is uniformly dense in its core regions.
\label{fig:sigma_1}}
\end{figure*}

Why, then, do the \squiggle post-starburst galaxies differ from other populations of post-starburst galaxies whether locally \citep{Wu2021} or at intermediate-redshift \citep{Wu2018, DeugenioF2020} which do show evidence of centrally concentrated star formation? The answer may lie in selection. High mass galaxies in samples selected using \hdelta equivalent width techniques like those in \cite{Wu2018} have been shown to have very small burst mass fractions, on the order of $\sim5\%$ \citep{French2018a}. The bursts in these galaxies are weak relative to the $>20\%$ bursts in \squiggle galaxies which are completely dominated by A-type populations \citep{Suess2022}. The galaxies with small bursts may represent a dusting of central star formation on top of an already quiescent population (which is similar to the toy model which does a poor job of describing \squiggle post-starburst size evolution) rather than a true quenching of a galaxy's primary epoch of star formation that occurs in the entire galaxy simultaneously. High redshift, massive post-starburst galaxies have been found to lack color gradients, suggesting that galaxies must shut off star-formation such that a young stellar population dominates the light at all radii on kiloparsec scales in the immediate aftermath of quenching \citep{Maltby2018, Suess2020}. These fundamental differences in the galaxy masses, burst mass fractions, and stellar age distributions between low- and high-redshift post-starburst galaxies may be the result of fundamentally different physical processes (e.g. increased major merger rates or higher gas fractions at earlier cosmic time). We suggest that the \squiggle post-starburst galaxies are the lower-z extension of that population of rapidly quenching galaxies, and explore in the following section what progenitors could have resulted in the formation of compact post-starburst galaxies without purely central star formation.

\subsection{Preferential fast quenching in compact star-forming galaxies}

The commonly invoked central starburst pathway to quiescence is often proposed as a way to take an extended star-forming galaxy and to produce a more concentrated elliptical galaxy. However, galaxies need not shrink down in the size-mass plane. Instead, these galaxies may have small sizes because they are the evolutionary descendants of more compact star-forming galaxies, which quenched rapidly after reaching a stellar density threshold that is correlated with other feedback mechanisms \citep[e.g.][]{VanDokkum2015}.

In Figure \ref{fig:sigma_1}a we show the relationship between the stellar mass and stellar surface mass density in the central kpc for the galaxies in \squiggle and LEGA-C. Post-starburst galaxies are only slightly denser than the quiescent galaxies at this redshift (median $\mathrm{log (\Sigma_{1 kpc, PSB}) - log (\Sigma_{1 kpc, Q})} = 0.10$). In contrast, the post-starburst and quiescent samples are significantly more dense than star-forming galaxies at fixed stellar mass (median $\mathrm{log (\Sigma_{1 kpc, PSB}) - log (\Sigma_{1 kpc, SF})} = 0.48$). The density of the post-starburst galaxies is similar to those found in previous studies of quiescent galaxies at this stellar mass, $\mathrm{log(\Sigma_{1 kpc})} \sim 10$ \citep[e.g.][]{Fang2013,VanDokkum2015,Barro2017,Mosleh2017,Whitaker2017,Suess2021}. This supports the finding that galaxy structure in the central regions is largely set at the time quenching occurs. In addition, we find that dense central structures are in place for the entirety of the \squiggle sample, regardless of the presence of tidal features; in Figure \ref{fig:sigma_1}b, we show that there is no significant difference in \sigone between the disturbed and non-disturbed populations. This does not preclude mergers helping the galaxies reach the central density threshold required for shutdown; simulations which rely on AGN feedback to quench galaxies have shown that only mergers which push galaxies into the regime where their central black holes are massive enough to trigger feedback effectively quench galaxies \citep{Quai2021}, and more rapid quenching is strongly associated with an increased injection of AGN feedback prior to quenching \citep{Park2021}. However, it does suggest that mergers may not be the smoking gun progenitor of all post-starburst galaxies. 

Because the central mass of a galaxy and the black hole mass are strongly correlated \citep{Kormendy2013}, it is natural to investigate the incidence of AGN in these dense post-starburst galaxies. The youngest galaxies in \squiggle host optical AGN at significantly higher rates than the oldest post-starburst galaxies and older quiescent galaxies \citep{Greene2020}. If feedback from AGN is connected to density, then it is possible that these galaxies went through a quasar phase which is still turning off in the youngest galaxies in the sample and that the galaxies will remain quenched after the AGN runs out of fuel and shuts down due to radio-mode feedback \citep[e.g.][]{Croton2006}. After that, minor mergers will contribute to the growth in outer density to grow galaxies without substantially changing their central density \citep[e.g.][]{Bezanson2009, VanDokkum2010}. This ex-situ growth could lead to a better match in the half-light radii of the galaxies as they grow their sizes without significantly changing their central structures. 

\begin{figure}
\includegraphics[width=0.45\textwidth]{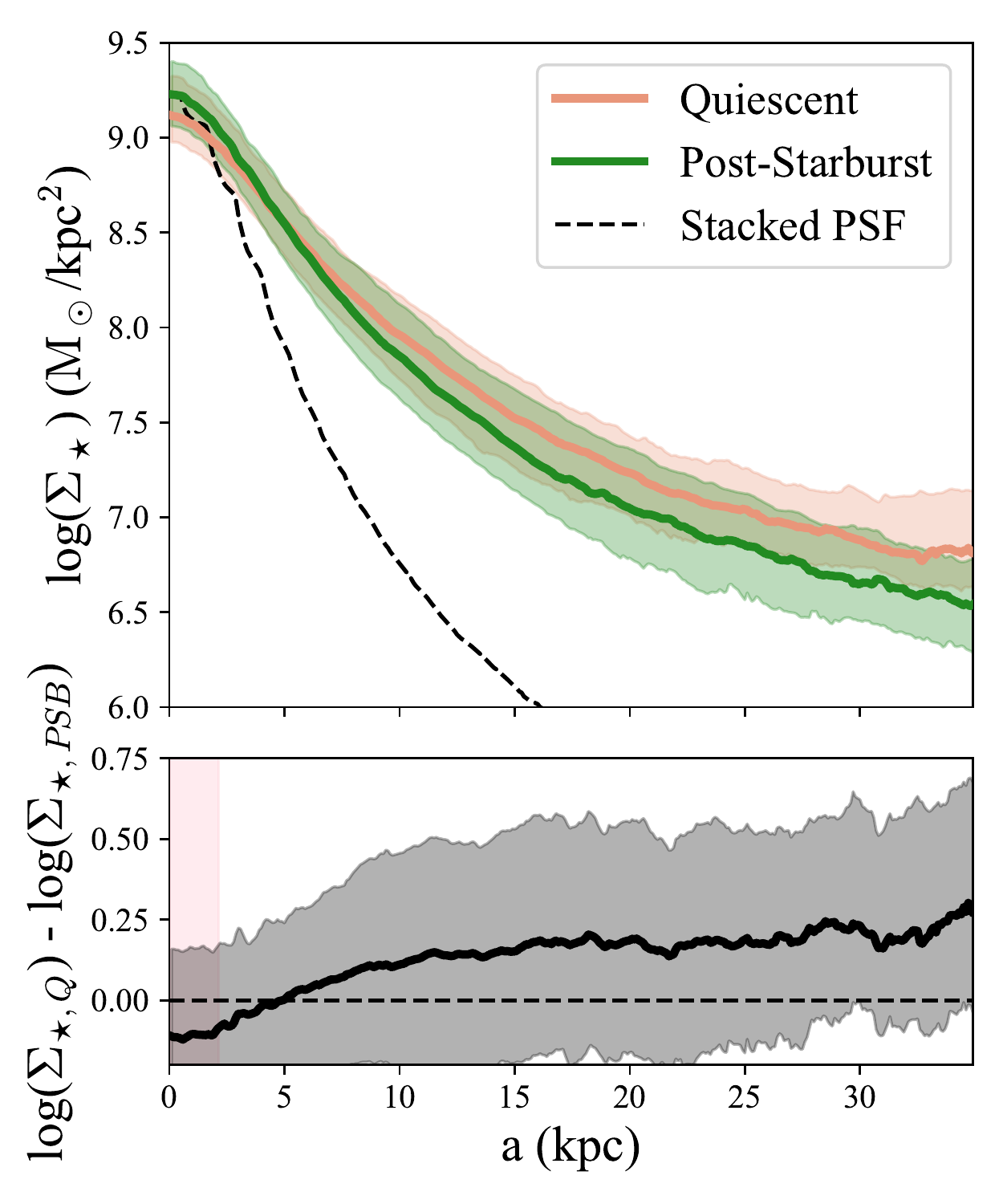}
\caption{One-dimensional median stacks of the surface mass profiles of \squiggle post-starburst galaxies (green) and the mass-matched LEGA-C control sample (red) as a function of the semi-major axis of the best-fitting ellipse in kiloparsecs. The profiles are truncated at the radius where the number of galaxies used in the stack drops below 20. Also shown is a stack of the model point-spread functions used in the fits (black dashed line). The bands bound the 16th and 84th percentile surface mass profiles for each sample. In the bottom panel, we show the difference between the quiescent and post-starburst median profiles as a black line with combined scatter indicated as a grey band, with the half-width half maximum seeing (0.3"$\sim$2.1 kpc at z=0.7) indicated by the pink shaded region. The differences in the profiles increases as a function of radius, flattening at $\sim10$ kpc.
\label{fig:surface_mass_comp}}
\end{figure}

The similar central densities and discrepant half-light radii of post-starburst and quiescent galaxies implies that the profiles should differ in shape most significantly in the wings if ex-situ growth is the dominant mode of evolution post-quenching. To investigate this, we derive one-dimensional surface mass profiles for the post-starburst and quiescent galaxies by multiplying the observed surface brightness profiles by the mass-to-light ratio, assuming no radial gradients. In Figure \ref{fig:surface_mass_comp}, we show the median surface mass profiles for the post-starburst and quiescent samples with associated scatter, as well as the difference between the median profiles. The ex-situ growth hypothesis is supported by these stellar-mass profiles, as the differences between the median profiles increases steadily out to $a=10$ kpc. This profile difference at large radii is likely less extreme in reality due to the empirical color gradients found in quiescent galaxies \citep[e.g.][]{Suess2019a, Suess2019b, Suess2020,Kawinwanichakij2021}. Quiescent galaxies tend to be bluer in their outskirts, which would lead to a smaller mass-to-light ratio and would alleviate at least some of the tension between these profile shapes. However, the color gradients present in quiescent samples are likely the result of the minor mergers we suggest are the dominant growth path post-quenching, and this normalization scheme still shows that the shapes of the profiles are most significantly different at $a>10$ kpc. Taken together, we suggest that in rapid quenching, the seeds of the structure of the quiescent population are formed in the recent burst on kpc scales, after which feedback correlated with density suppresses any additional star formation and minor mergers become the primary form of galaxy mass growth. 

One possible class of progenitors could be dusty, extreme starbursts, or ``sub-mm galaxies" \citep{Toft2012,Toft2014, Wild2020}, which have the extremely high star formation rates and feedback necessary to produce galaxies like those in \squiggle \citep{Spilker2020a, Spilker2020b}. At high redshift, sub-mm galaxies also lie slightly below quiescent galaxies in the mass-size plane \citep{Gomez-Guijarro2022}. If they shut off their rapid star formation abruptly and uniformly, sub-mm galaxies would likely result in post-starburst SEDs and flat age gradients similar to those measured in \squiggle galaxies \citep{Setton2020}. While detailed analysis of number densities is beyond the scope of this work, we note that in LEGA-C, star-forming galaxies do exist which are dense enough in $\mathrm{log(\Sigma_{1 kpc})}$ to evolve into \squiggle post-starburst galaxies without a significant amount of additional compaction. These galaxies do not have star formation rates high enough to be sub-mm galaxies, but it is unlikely that such extremely rare progenitors would be found in a field as small as that of LEGA-C. Future work will focus on more strongly constraining the density of \squigglecomma-like post-starburst galaxies as a function of redshift to tie them more concretely to a progenitor population.

Ultimately, we conclude that the recent star formation in post-starburst galaxies must have taken place on spatial scales which are comparable to the size of the galaxy, as we do not see any evidence of the type of fading that would be expected from a centralized starburst. Although galaxy mergers likely play a role in quenching some of the post-starburst galaxies in \squiggle (see Verrico et al. in preparation), the finding that log(\sigonecomma) is fully consistent between the disturbed and non-disturbed samples and the lack of positive size evolution suggests that merger driven central starbursts are not the entire story. Instead, the mergers may, in some galaxies, simply be the last push towards a central density high enough to trigger feedback which rapidly shuts off star formation while locking in existing structure and maintains the shutdown. After this, galaxies could passively evolve into red, quiescent galaxies that grow in half-light size via minor mergers but otherwise do not significantly change in their structures.

\section{Conclusions}

Using images from the Hyper-Suprime Cam Survey, we study the sizes and structures of $z\sim0.7$ post starburst galaxies from the \squiggle Survey \citep{Suess2022} in comparison to coeval massive galaxies in the LEGA-C Survey. By performing single-component \sersic fitting of the galaxies, we have robustly measured sizes which account for non-\sersic low surface brightness features and structural measures like the \sersic index and the axis ratio. In addition, we measure non-parametric indicators of concentration. We conclude the following:

\begin{itemize}

    \item Post-starburst galaxies have smaller half-light radii than coeval star-forming and quiescent galaxies at similar stellar mass (see Figures \ref{fig:mass_size} and \ref{fig:delta_re_hists}). Specifically, they are systematically $\sim$0.1 dex smaller than quiescent galaxies and $\sim0.4$ dex smaller than star-forming galaxies. The compactness at fixed stellar mass does not vary strongly based on whether or not a galaxy is tidally disturbed.
    
    \item The sizes and structures of post-starburst galaxies, as measured via parametric (see Figure \ref{fig:n_ar_hists}) or non-parametric measures (see Figure \ref{fig:gini_gc}) also point to concentrated, round galaxies which are more similar to quiescent galaxies than they are to star-forming galaxies.
    
    \item Post-starburst galaxies either negative or flat correlation between their sizes and their time since quenching (see Figure \ref{fig:re_tq}). This trend stands in constrast with what would be expected for a fading central starburst, indicating that the recent burst of star formation was not limited to the galaxy center, and instead must have occurred on larger ($\gtrapprox1$ kpc) spatial scales.
    
    \item The central densities of post-starburst galaxies are very similar to those of quiescent galaxies (see Figure \ref{fig:sigma_1}) and match the common threshold for quiescence, log(\sigonecomma) $\sim10$ found in the literature. 
    
    \item The median shape of the post-starburst and quiescent surface brightness profiles are most significantly different at large radius (a$>$10 kpc), where the quiescent sample has more light (see Figure \ref{fig:surface_mass_comp}). This indicates that while the central shapes of galaxies do not change significantly once quiescence is reached, the outer envelopes may indeed grow via minor merging which deposits stellar mass as large radius. 
    
    
\end{itemize}

Two competing forces are at play in the detailed study of the rapid mode of quenching. On the one hand, fast quenching dominates at high redshift, so in the ideal case one would hunt for post-starburst galaxies at the highest redshift possible in order to identify candidates for galaxies which have recently shut off their primary epoch of star formation. On the other hand, in order to push to high redshift, deep integration times are required, which makes it difficult to cover the large patches of sky necessary to identify galaxies in this short-lived period of evolution. The ideal situation (deep, red spectroscopic surveys of large areas of the sky), however, is right around the corner with the impending public releases of deep spectroscopic surveys like the Dark Energy Spectroscopic Instrument \citep{DESI} and the Prime Focus Spectrograph \citep{Takada2014} surveys. In the very near future, the high-quality spectra from these surveys will allow for the identification of large samples of post-starburst galaxies out past $z\sim 1$. Future studies of the galaxies identified in these surveys will significantly bolster our understanding of the rapid mode of quenching by identifying the first statistically large samples of post-starburst galaxies near the era of cosmic noon. The quantification of the number density, size, structure, AGN activity, and gas content of post-starburst galaxies as a function of redshift will place strong constraints on the rapid quenching pathway of galaxy evolution.

\acknowledgements 

DS, JEG, RSB, and DN gratefully acknowledge support from NSF grant AST1907723. This work was performed in part at the Aspen Center for Physics, which is supported by National Science Foundation grant PHY-1607611. DS would also like to thank the North American ALMA Science center for financial support in attending the ACP ``Quenching" conference. MV acknowledges support from the NASA Pennsylvania Space Grant Consortium. RF acknowledges financial support from the Swiss National Science Foundation (grant no 157591 and 194814). DS thanks Alan Pearl, Brett Andrews, Yasha Kaushal, and Alwin Mao for their help with visual classifications as well as in perfecting colormaps and point sizes down to the minutia. Finally, DS thanks Stephanie Permut for her kindness in listening to ramblings about ``intermediate-redshift post-starburst galaxies" for months on end. 

The Hyper Suprime-Cam (HSC) collaboration includes the astronomical communities of Japan and Taiwan, and Princeton University. The HSC instrumentation and software were developed by the National Astronomical Observatory of Japan (NAOJ), the Kavli Institute for the Physics and Mathematics of the Universe (Kavli IPMU), the University of Tokyo, the High Energy
Accelerator Research Organization (KEK), the Academia Sinica Institute for Astronomy and Astrophysics in Taiwan (ASIAA), and Princeton University. 

Funding for SDSS-III has been provided by the Alfred P. Sloan Foundation, the Participating Institutions, the National Science Foundation, and the U.S. Department of Energy Office of Science. The SDSS-III web site is http://www.sdss3.org/. 

SDSS-III is managed by the Astrophysical Research Consortium for the Participating Institutions of the SDSS-III Collaboration including the University of Arizona, the Brazilian Participation Group, Brookhaven National Laboratory, Carnegie Mellon University, University of Florida, the French Participation Group, the German Participation Group, Harvard University, the Instituto de Astrofisica de Canarias, the Michigan State/Notre Dame/JINA Participation Group, Johns Hopkins University, Lawrence Berkeley National Laboratory, Max Planck Institute for Astrophysics, Max Planck Institute for Extraterrestrial Physics, New Mexico State University, New York University, Ohio State University, Pennsylvania State University, University of Portsmouth, Princeton University, the Spanish Participation Group, University of Tokyo, University of Utah, Vanderbilt University, University of Virginia, University of Washington, and Yale University.

\newpage

\bibliography{all}

\appendix 

\section{Verifying the accuracy of structural measurements using ground-based imaging  \label{sec:hst_comp}}

The i band galaxy images from the HSC survey are remarkable both for their depths and resolutions from the ground, but their resolution (PSF FWHM $\sim0.6$") is still dwarfed by that of space based instruments like the Hubble Space Telescope (PSF FWHM $\sim0.15$"). We have turned to the HSC survey because obtaining space based images of the entire \squiggle post-starburst sample is unfeasible; the galaxies are distributed throughout the entire SDSS footprint and it would require a significant investment to get HST followup imaging of the entire sample. However, the LEGA-C Survey, which we utilize as a coeval comparison sample of galaxies, completely overlaps with HSC and also has existing HST/ACS F814 imaging. This means that we can directly compare the \sersic sizes and structures we derive for LEGA-C galaxies to those measured in \cite{VanDerWel2021} to test how our ground based measurements compare to those from space.

\begin{figure*}
\includegraphics[width=\textwidth]{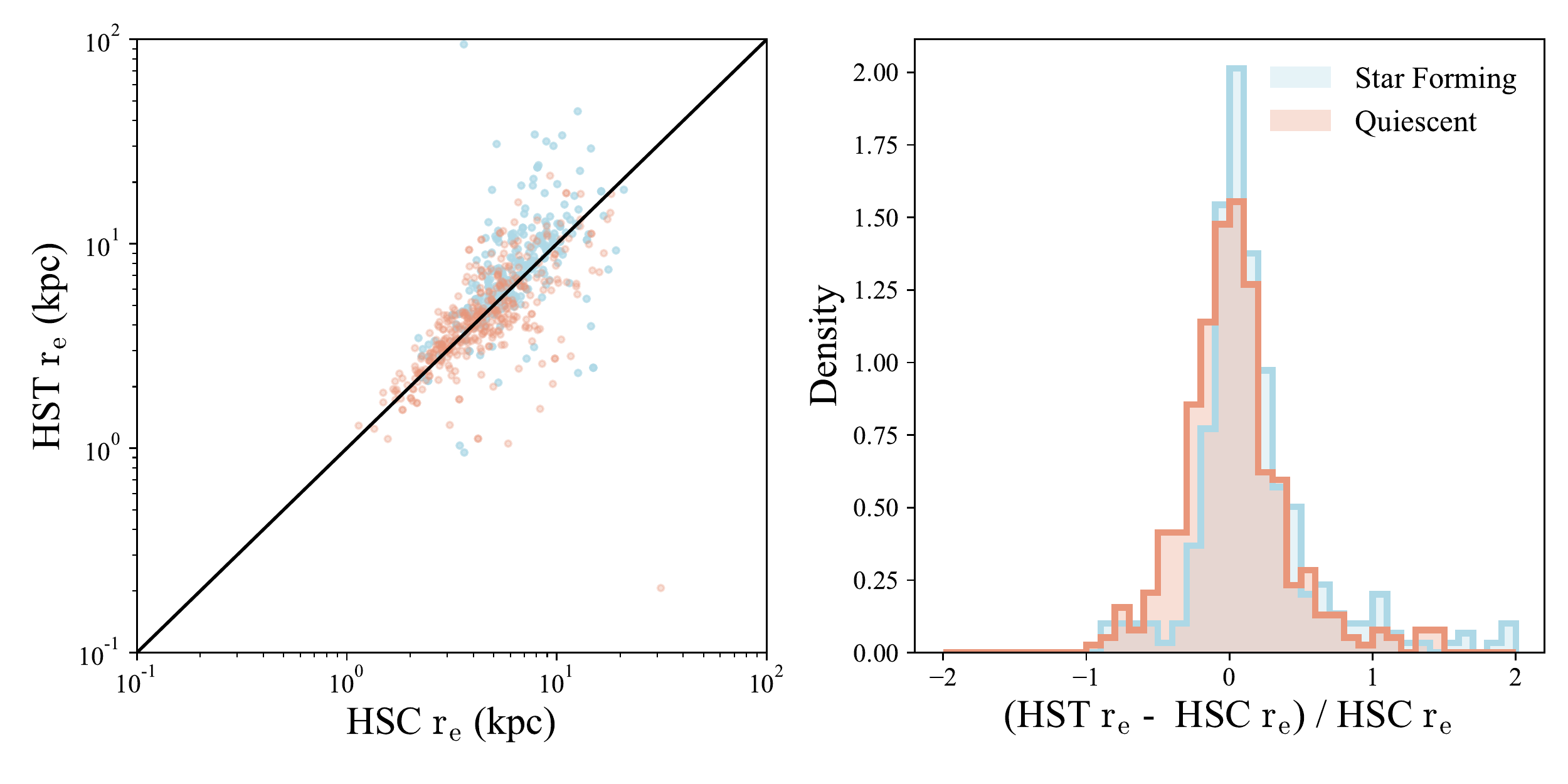}
\caption{(Left) The one-to-one relation between the \sersic effective radii we measure on the HSC images and the \sersic radii from \cite{VanDerWel2021} measured from HST/ACS images. Star forming galaxies are colored as blue, and quiescent galaxies are red. (Right) The percent difference in the measurements for the two LEGA-C samples. For quiescent galaxies, the median sizes we measure are extremely close, with a scatter of $\sim20\%$. For star forming galaxies, the scatter is similar, but there is a systematic offset where the HST sizes are $\sim8\%$ larger than the ones we measure. Both sub-samples have longer tails in the direction of larger HST sizes. 
\label{fig:hst_hsc_re_comp}}
\end{figure*}

\begin{figure*}
\includegraphics[width=\textwidth]{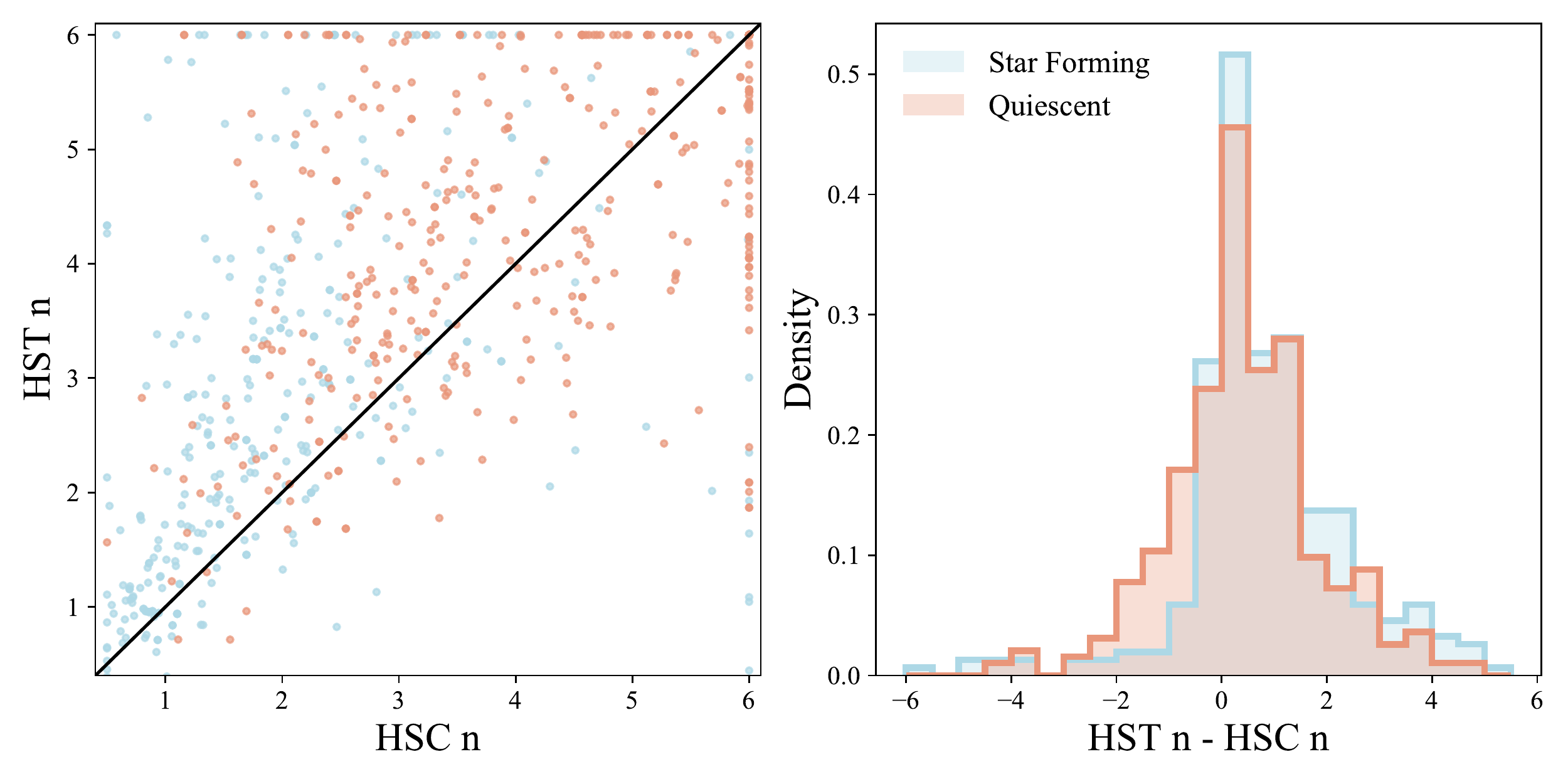}
\caption{As in Figure \ref{fig:hst_hsc_re_comp}, but for the \sersic indices. In both the quiescent and star forming subsamples, the \sersic indices measured in HST tend to be larger than those we measure with HST, but the trend is more pronounced for the star forming galaxies. 
\label{fig:hst_hsc_n_comp}}
\end{figure*}

In Figure \ref{fig:hst_hsc_re_comp} and Figure \ref{fig:hst_hsc_n_comp}, we show the comparisons between our measured sizes and \sersic indices and the values measured from the HST images, split into quiescent (red) and star-forming (blue) populations. For the majority of galaxies, the sizes are fairly well recovered, especially in the quiescent populations where the percent error in the sizes scatters around $0\%$. However, for star-forming galaxies, there is a systematic offset, where the galaxies are measured in HST to be a median of $\sim8\%$ larger than they are in HSC imaging. This difference is reflected in the \sersic indices as well; quiescent populations are measured in HST with slightly higher \sersic indices, but the difference is more pronounced in the star-forming populations. 

\begin{figure*}
\includegraphics[width=\textwidth]{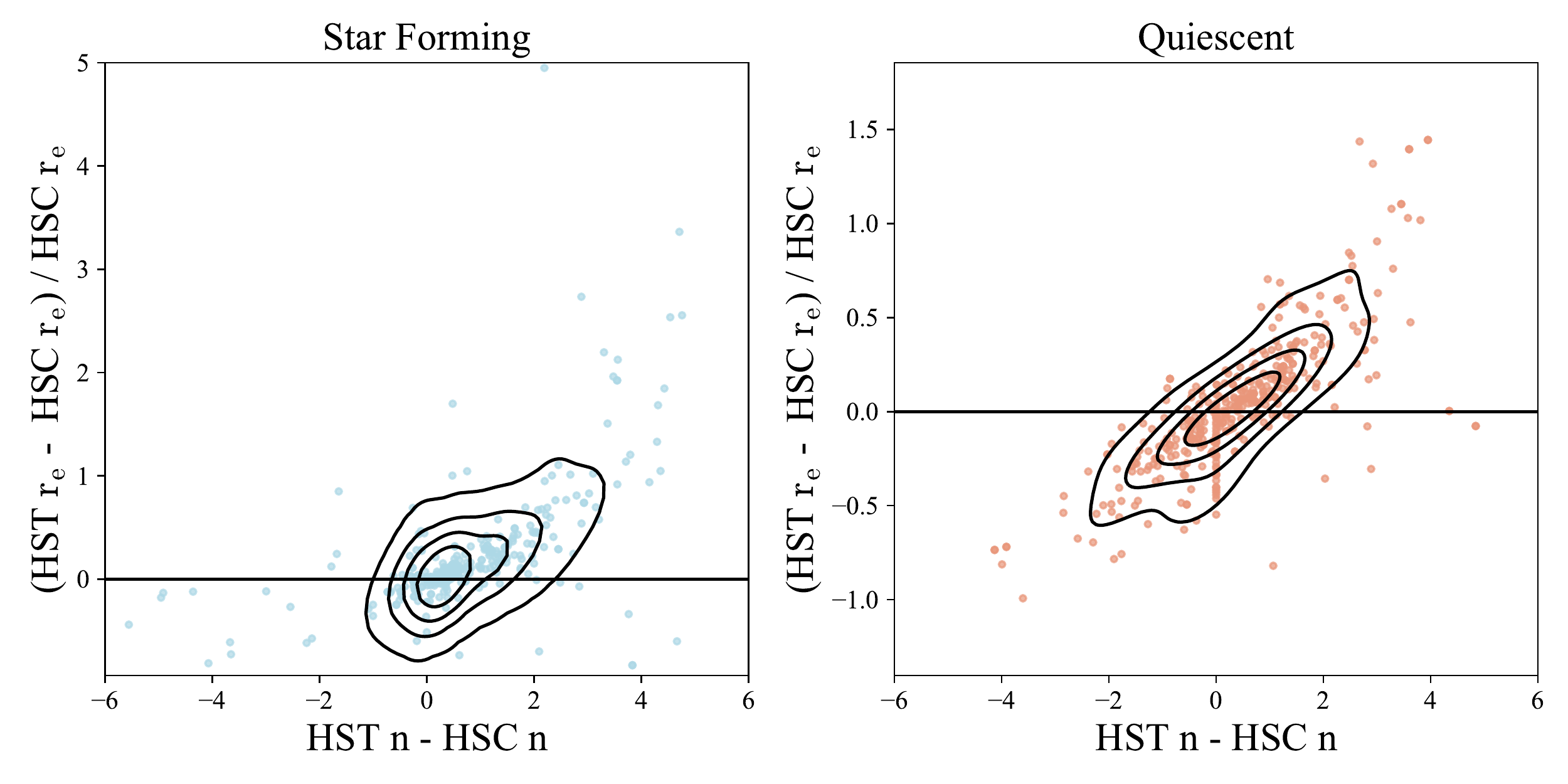}
\caption{The percent difference in the size vs the difference in \sersic index for the star forming (left) and quiescent (right) sub-samples, with contours bounding 80\% of the galaxies. There is a trend in both samples where a mismatch in \sersic index correlates with a mismatch in the measured sizes. (Note: We truncate the y-axis on the star forming plot at 5 despite a single point at $\sim25$ due to a non-physical HST measurement of that galaxy.)
\label{fig:delta_delta}}
\end{figure*}

\begin{figure*}
\includegraphics[width=\textwidth]{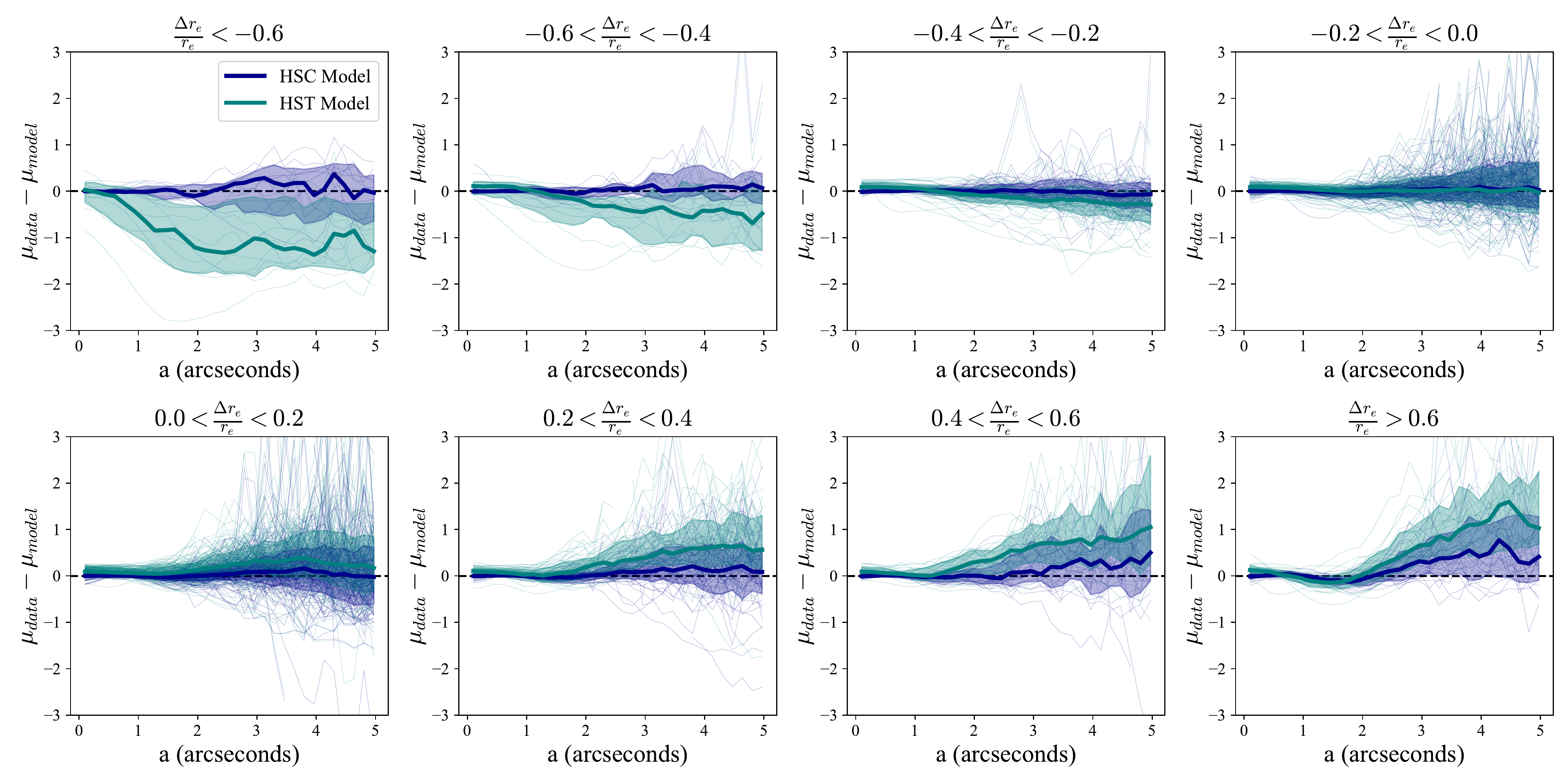}
\caption{Individual (thin lines) and median (thick lines) residuals between the surface brightness profiles extracted from the cutouts and the best fitting models from this work (dark blue) and the fits to HST ACS images described in \cite{VanDerWel2016} convolved with the HSC PSF (teal). Separate panels show the galaxies binned by the agreement between the two fits in the effective radius. Profiles are only shown for galaxies where no nearby objects are simultaneously fit. In all cases, our model surface brightness profiles are consistent with no residuals. However, the HST fits which overestimate the sizes do so because they have significantly more light in the wings than our galaxies, as well as slightly more peaked cores.
\label{fig:delta_mu_re}}
\end{figure*}

\begin{figure*}
\includegraphics[width=\textwidth]{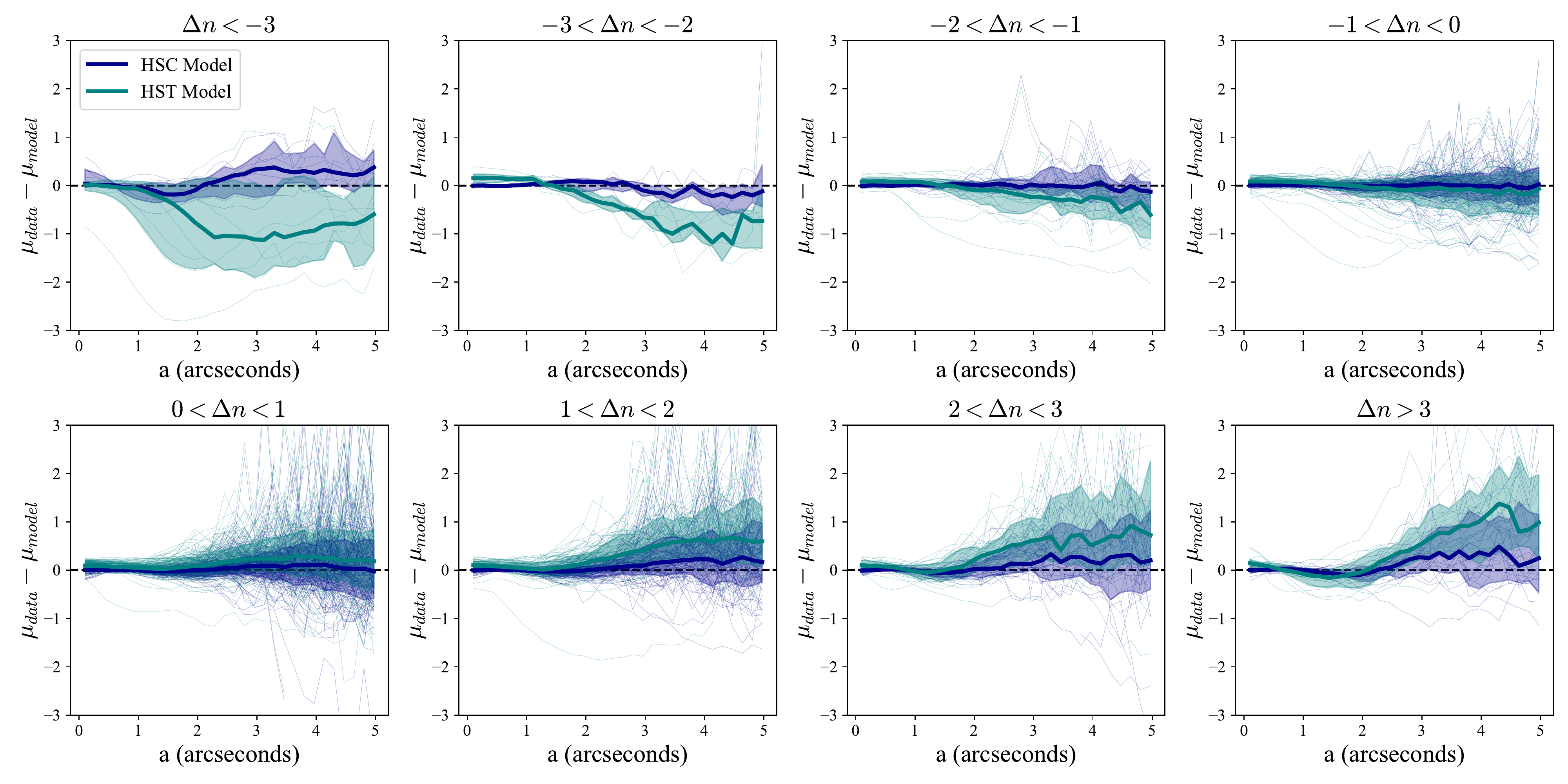}
\caption{As in Figure \ref{fig:delta_mu_re}, but this time binned by the agreement between the \sersic indices of the fits. As with the sizes, the differences in \sersic indices result from significant failures to successfully fit the low surface brightness wings in the HST images.
\label{fig:delta_mu_n}}
\end{figure*}

In Figure \ref{fig:delta_delta}, we show the covariance between these two offsets. Galaxies which are fit in HST with larger sizes are also fit with larger \sersic indices, and this tail at the larger-size/higher-\sersic index end is more pronounced in the star-forming population. Because high \sersic index corresponds to both a more peaked core and more extended wings, we turn to 1D surface brightness profiles to understand what is driving this difference. In order to do so, we generate models of the best fitting space-based galaxy parameters from \cite{VanDerWel2021} at the pixel scale of HSC and convolved with the same HSC PSF we use to fit the galaxies. In order to avoid systematic differences in source identification and deblending, we restrict this test to galaxies which do not have a bright neighbor within 25 pixels. In Figure \ref{fig:delta_mu_re} and Figure \ref{fig:delta_mu_n}, we show residuals in the surface brightness profiles for galaxies as fit in this work (dark blue) and with the best fitting models fit to the HST images of the galaxies under HSC viewing conditions (teal), binned by the offset between the two measurements. We find that the differences in size and \sersic index are largely driven by a difference in the wings. The HST \sersic structures overpredict the amount of light that will be present at large radius, and, as a result, measure larger sizes. This is likely due to the sensitivity of HST to the cores of the galaxies; in trying to accurately fit the peaky centers of the galaxies, the HST fits converge to large n, which is compensated for by inflated sky values. In contrast, HSC's remarkable low-surface brightness limits ($\sim28.5$ $\mathrm{mag/arcsecond^2}$) allows for well calibrated sky measurements which result in small residuals at large radius. This better allows for the galaxies to be fit with low \sersic indices, which is more significant in the star-forming galaxies which tend to be more disk-like.

\begin{figure*}
\includegraphics[width=\textwidth]{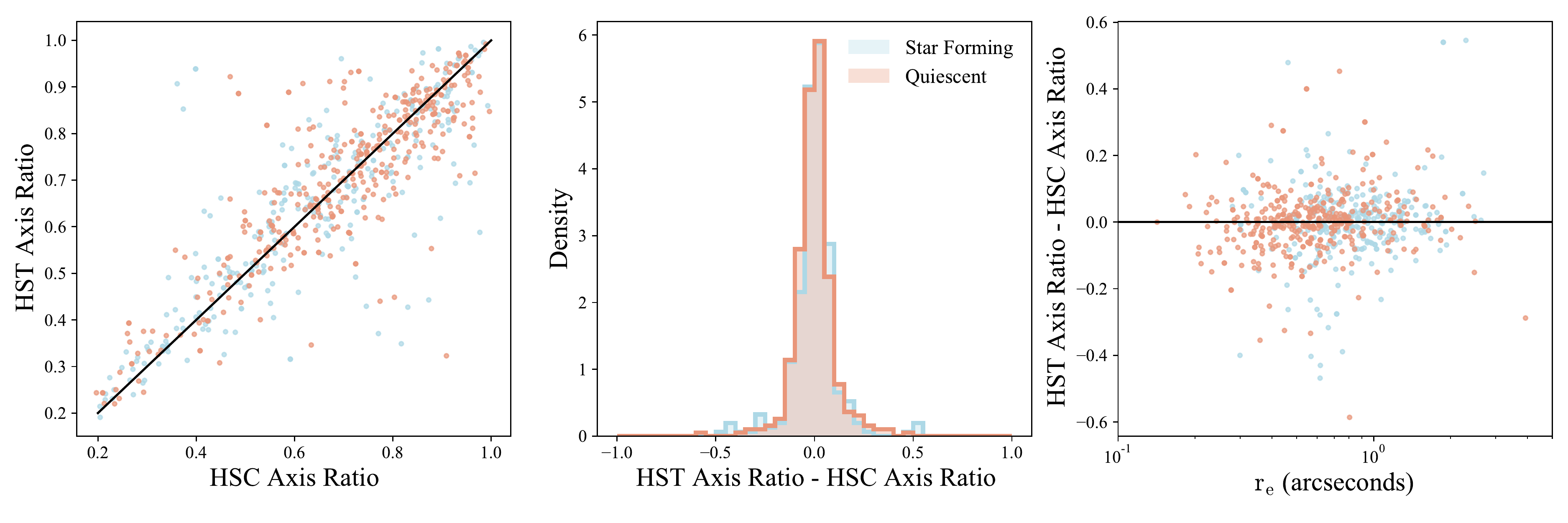}
\caption{(Right) The axis ratio we measure using HSC data for quiescent (red) and star-forming (blue) LEGA-C galaxies versus the same parameter measured using HST data. A 1:1 correspondence is shown in black. (Center) The difference in the measurements of the axis ratios. The median deviation is consistent with zero and the scatter is very small. (Right) The difference between the axis ratios as a function of the size in arcseconds. The smallest galaxies are as well recovered as the largest ones, indicating that even the smallest galaxies in \squiggle which are resolved have axis ratio measurements that can be trusted. 
\label{fig:ar_comp}}
\end{figure*}

Our fitting does an especially good job at recovering the shapes of galaxies in their axis ratios. We illustrate this in Figure \ref{fig:ar_comp}, showing the 1:1 relation between the projected axis ratios we measure and those from \cite{VanDerWel2016}. The values agree between the ground and space based measurements with a scatter of only $\sim0.07$, indicating that even with significantly more PSF smearing, the projected galaxy shapes are still recoverable using \texttt{GALFIT}. Perhaps more important, there does not appear to be any trend with the sizes of the galaxies; in the right panel of Figure \ref{fig:ar_comp}, we show that the difference in axis ratios does not correlate with size for either the star-forming or quiescent control samples. Even at the low size (\re$\sim$0.1") limit where the sizes of the galaxies are significantly smaller than the PSF, the axis ratios are still robustly recovered. This indicates that the roundness that we measure in \squiggle galaxies is not just a resolution effect, but a real physical property we can trust in our interpretations. 

We conclude that ground-based imaging with low surface brightness limits is extremely well suited for the task of measuring galaxies sizes and \sersic structures. While the loss in resolution from HST to HSC does not allow for the centers of galaxies to be resolved, the increase in sensitivity at large radius allows for the total light of the galaxy to be better accounted for, which improves the measurement of \re. In addition, we find that observing from the ground does not affect measurements of the axis ratio significantly, even for the smallest galaxies. 

\section{Comparing the smallest galaxies to an unresolved point source \label{sec:resolved_check}}

Despite the agreement between the sizes fit on the LEGA-C control sample and those in HST, resolution is still a concern. The smallest galaxies in the \squiggle sample are fit with sizes $\sim$0.1", whereas the median seeing in the HSC i-band is $\sim0.6$". In order to confirm that all the sizes we report for galaxies are reliable from a ground based survey, we re-run all our fits using an identical algorithm to the one outlined in Section \ref{subsec:sersic}, but this time we force the two-dimensional galaxy profiles to be a point source convolved with the PSF we provide to \texttt{GALFIT} from the HSC PSF Picker. We calculate radial $\chi^2$ values both the \sersic and PSF-only models out to the point where the signal-to-noise in the galaxy profile drops below 5, and find that the PSF-only models have median $\chi^2$ values $\sim$2.5 orders of magnitude higher than for the \sersic models convolved with the PSF, indicating that the \squiggle sample is significantly resolved. Even the best performing PSF models are still significantly worse fits than \sersic profiles.

\begin{figure*}
\centering
\includegraphics[width=0.92\textwidth]{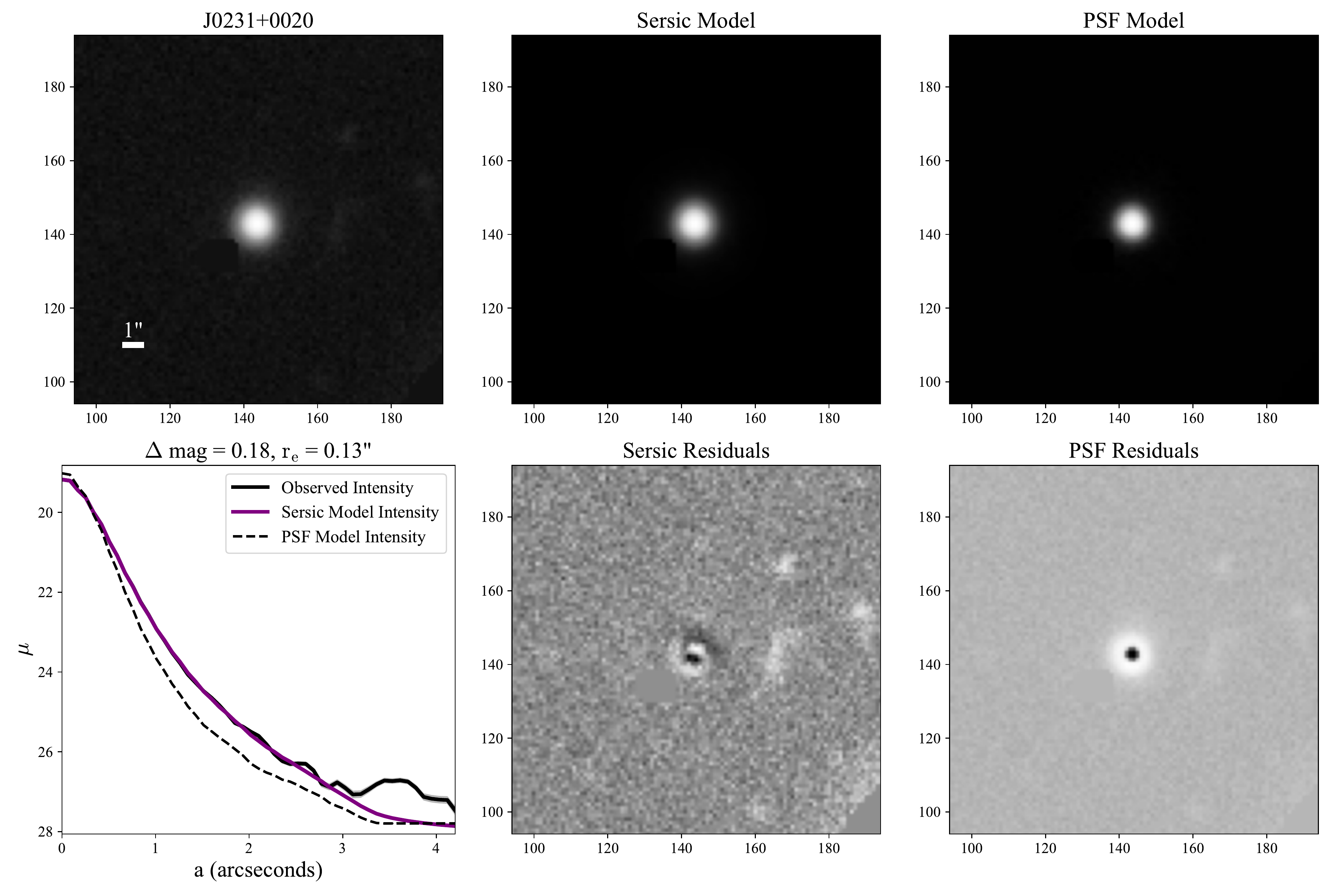}
\includegraphics[width=0.92\textwidth]{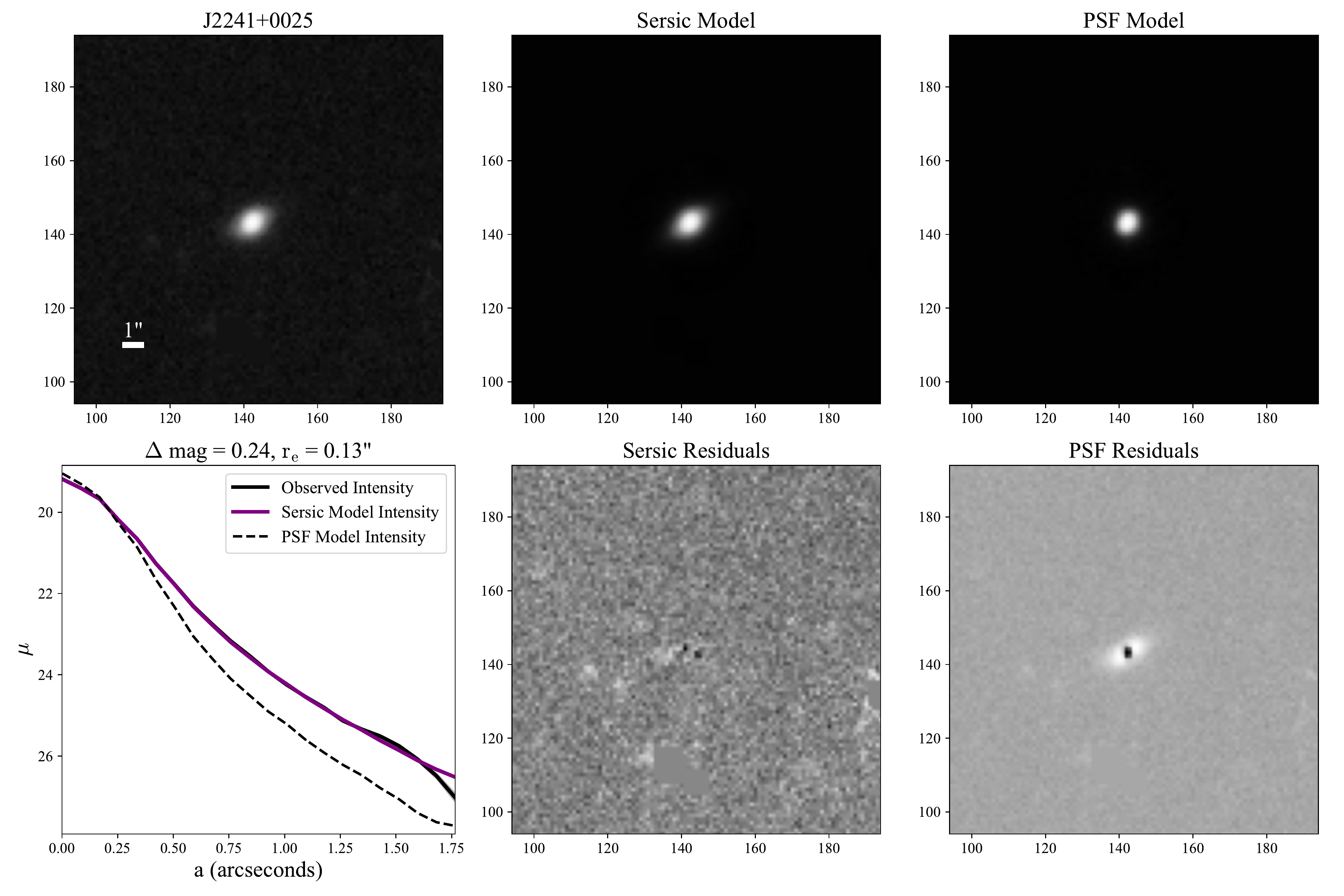}

\caption{In the top row, $\sim$17x17" cutouts of the image (left), best fitting \sersic model (center), and best fitting point source model (right) generated with \texttt{GALFIT} for the two smallest galaxies in \squiggle with no neighbors which required simultaneous fitting. Below each of the models, we show the residuals, with masked pixels greyed out. In the bottom left, we show the 1D surface brightness profiles for the data (blue), the \sersic model (red), and the point source (black). Profiles are truncated when the signal to noise drops below 3. It is clear that even for \re$\sim0.1$" galaxies, the best fitting point source model will result in a galaxy which is too centrally peaked and which does not properly capture the galaxy wings seen in the data. 
\label{fig:sersic_psf_comp}}
\end{figure*}

To illustrate this, in Figure \ref{fig:sersic_psf_comp} we show the two smallest galaxies in \squiggle which do not have bright nearby companions, as well as the 2D and 1D best fitting \sersic and point source models. The 1D profiles are extracted using the best fitting axis ratio from the \sersic model. It is clear from the residuals and the 1D profile that the \sersic model better captures the true shape of the galaxy profiles, both in the centers (where the PSF is far too peaked) and in the wings. In addition, in the case of J2241+0025, there is a clear elongation in the galaxy that that the PSF shape cannot account for. Thus, even though the galaxy sizes we fit are small, we still consider them to be reliable when the PSF is properly accounted for in the fits. 

\end{document}